\documentclass[aps,pra,reprint]{revtex4-1}

\usepackage{xcolor}
\usepackage{tikz}
\usepackage{amsmath}
\usepackage{amssymb,amsmath}
\usepackage{graphicx}
\usepackage{upgreek}
\usepackage[T1]{fontenc}
\usepackage[utf8]{inputenc}
\usepackage{graphicx}
\usepackage{bm}

\begin{document}


\title{Kerker Effect, Superscattering, and Scattering Dark State in Atomic Antennas}

\author{Rasoul Alaee{$^{*1,2}$}, Akbar Safari$^{1}$, Vahid Sandoghdar{$^{2}$}, and Robert W. Boyd{$^{1}$}}
\address{
{$^{1}$}Department of Physics, University of Ottawa, 25 Templeton, Ottawa, Ontario, K1N 6N5, Canada\\
{$^{2}$}Max Planck Institute for the Science of Light, Erlangen 91058, Germany\\
$*$ $\rm{Corresponding\,\, author: rasoul.alaee@gmail.com}$
}


\begin{abstract}
We study scattering phenomena such as the Kerker effect, superscattering, and scattering dark states in a subwavelength atomic antenna consisting of atoms with only electric dipole transitions. We show that an atomic antenna can exhibit arbitrarily large or small scattering cross sections depending on the geometry of the structure and the direction of the impinging light. We also demonstrate that atoms with only an electric dipole transition can exhibit a directional radiation pattern with zero backscattering when placed in a certain configuration. This is a special case of a phenomenon known as the Kerker effect, which typically occurs in the presence of both electric and magnetic transitions. Our findings open a pathway to design highly directional emitters, nonradiating sources, and highly scattering objects based on individually controlled atoms. \end{abstract}

                          
\maketitle
Scattering is a fundamental phenomenon in light–matter interaction, and the scattering cross section is a measure to describe how strong an object interacts with incident photons~\cite{Bohren2008}. The ability to manipulate and control scattering is an important research goal in optics and atomic physics. In nanophotonics, the scattering can be controlled by engineering the geometry of the optical antennas as the canonical elements of the interaction \cite{Sandoghdar:Book2013}. In particular, it has been shown that engineered nanoparticles can exhibit remarkable scattering phenomena such as directional radiation patterns with zero backscattering known as the Kerker effect~\cite{kerker1983,Zambrana:2013,Alaee_kerker:15}, scattering dark state~\cite{Devaney:1973,Hsu:2014}, and superscattering~\cite{Ruan2010,Ruan2011,Qian:2019ExpSuperscattering}. To observe these phenomena, the nanoparticles are designed to support \textit{not only} electric dipole, but also magnetic dipole and higher-order multipoles. The underlying physics of these scattering phenomena can be understood from the interference among the scattered fields of all the induced multipole moments~\cite{kerker1983,Zambrana:2013,Alaee_kerker:15,Devaney:1973,Hsu:2014,Ruan2010,Alaee2019,Qian:2019ExpSuperscattering}.

\begin{figure}
    \centering
    \includegraphics[width=0.48\textwidth]{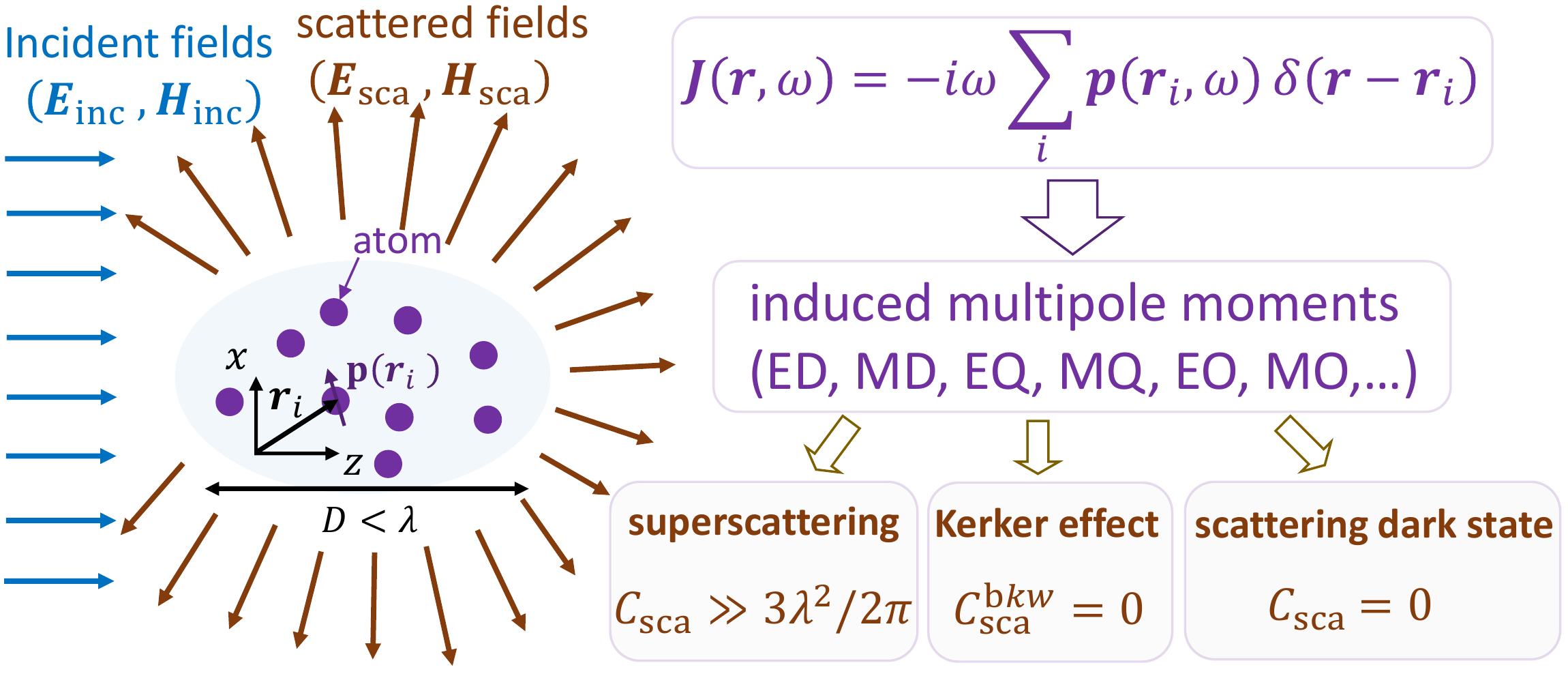}
    \caption{\textit{Main idea of this letter:} Sketch of a subwavelength atomic antenna composed of a few atoms with only \textit{electric} dipole transition moments. The overall size of the antenna is smaller than the wavelength i.e., $D<\lambda$. The induced displacement current $\mathbf{J}\left(\mathbf{r},\omega\right)$ of the antenna can be found from the induced dipole moment of the atoms, $\mathbf{p}\left(\mathbf{r}_{i}\right)$ placed at $\mathbf{r}_{i}$. The induced current is decomposed to multipole moments which are used to control the scattering cross section of the structure. The excitation and the position of the atoms can be tailored to achieve \textit{arbitrary} induced electric and magnetic multipoles. Thus, it is possible to control the scattering and achieve scattering dark states, superscattering, and the Kerker effect. The maximum scattering cross section of a single atom, $3\lambda^2/2\pi$, is considered as the measure.}
    \label{fig:MainIdea}
\end{figure}

With the recent developments in manipulating and controlling single atoms and quantum emitters, a new class of subwavelength antennas consisting of quantum emitters have been employed in the past decades to control light-matter interaction at the atomic scale~\cite{Jenkins:2012,Feng2013,Bettles2015,Bettles2016Cooperative,Bettles2016,Facchinetti2016storing,Asenjo2017,shahmoon2017,Barredo:2018,wild2018quantum,Facchinetti2018,bettles2019quantum,Zoller2019,Genes2019,grankin2018,mkhitaryan2018,Sandoghdar2020NL,Bekenstein2020,Rui:2020,Alaee2020,Browaeys2020,Ballantine2020}. Here, it should be borne in mind that atoms arranged in a subwavelength structure interact with the radiation field cooperatively~\cite{Dicke:1954,Lehmberg1970,Friedberg1973,Haroche1982}. Hence, although the canonical elements of these systems remain atoms with only electric dipole transitions, it turns out that the entire structure can be modeled as a single antenna that exhibits both electric and magnetic multipoles \textit{effectively}, and thus, enables the control of light scattering~(see Fig.~\ref{fig:MainIdea}). 

In this Letter, we treat the situation of a few atoms in simple subwavelength geometries and show that the scattering of the structure can be significantly different from the scattering of the individual atoms (see Fig.~\ref{fig:MainIdea}). Each atom supports only electric dipole transitions with the maximum scattering cross section of $3\lambda^2/2\pi$~\cite{lagendijk1996,Lagendijk1998,Zumofen:2008,Alaee:2017Review}. Using the multipole expansion~\cite{Alaee:2018,Alaee2019,SM}, the induced \textit{effective} electric and magnetic multipole moments of the atomic antenna can be found from the induced current $\mathbf{J}\left(\mathbf{r},\omega\right)$. By employing these multipole moments, we show in the following that it is possible to tailor the scattering of the structure to achieve extremely large or small scattering cross sections (the so-called superscattering and scattering dark state, respectively), as well as the Kerker effect. 

\begin{figure}
    \centering
    \includegraphics[width=0.48\textwidth]{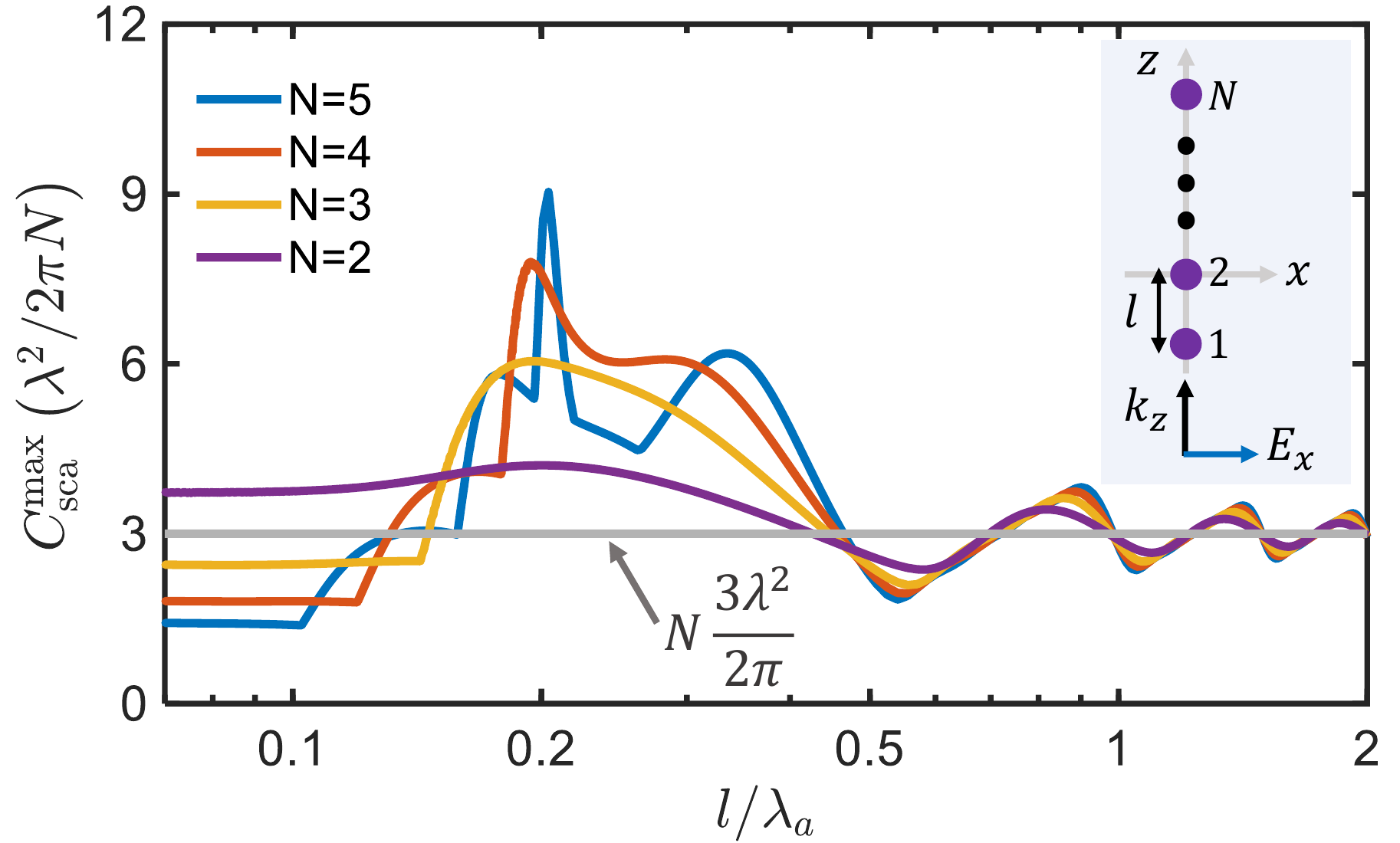}
    \caption{\textit{Superscattering: dimer, trimer, tetramer, and pentamer.} Maximum scattering cross section of $N$ atoms normalized to $\text{\ensuremath{\lambda^{2}/2\pi}}$ as a function of the distance between the atoms, $l$. The gray line shows the maximum scattering cross section for $N$ non-interacting atoms, i.e., $N\times\frac{3\lambda^{2}}{2\pi}$. The geometry of the atoms and the driving field is shown in the inset. The maximum scattering occurs at different frequency detuning for each $l$.}
    \label{fig:Superscattering_N_atoms}
\end{figure}
\textit{Atomic antenna in the weak excitation limit.—} The atoms in Fig.~\ref{fig:MainIdea} are modeled as two-level systems and are arranged such that the overall size of the antenna is smaller than the wavelength of the incident light. We assume an isotropic and linear atomic response, i.e., a weak excitation such that the atomic transition is far below the saturation limit. Therefore, the electric polarizability of each atom amounts to $\alpha\left(\omega\right)=-\left(\alpha_{0}\Gamma_{0}/2\right)/\left(\delta+i\Gamma_{0}/2\right)$, where, $\alpha_0=6\pi/k^{3}$, and $k$ is the wavenumber of the resonant light, $\Gamma_{0}$ is the radiative decay rate of the atomic transition at frequency $\omega_{a}$, and $\delta=\omega-\omega_{a}\ll\omega_{a}$ represents the frequency detuning between the illumination and the atomic resonance~\cite{Lagendijk1998,lambropoulos2007}. 
The antenna is illuminated by a plane wave $\mathbf{E}_{\mathrm{inc}}\left(\mathbf{r}\right)=E_{0}e^{i \mathbf{k}\cdot\mathbf{r}}\mathbf{e}_{E}$ propagating in the $\mathbf{k}$ direction, where $\mathbf{e}_{E}$ is the unit vector in the electric field direction, and $E_{0}$ is the electric field amplitude. The extinction cross section is defined as the ratio of the extracted power $P_{\mathrm{ext}}$ to the incident intensity $I_0$ and is given by
\begin{equation}
C_{\mathrm{ext}}=\frac{P_{\mathrm{ext}}}{I_{0}}=\frac{k}{\epsilon_{0}\left|E_{0}\right|^{2}}\sum\limits_{i=1}^{N}\mathrm{Im}\left[\mathbf{p}\left(\mathbf{r}_{i}\right)\cdot\mathbf{E}_{\mathrm{inc}}^{*}\left(\mathbf{r}_{i}\right)\right],\label{eq:Ext_def}
\end{equation}
where $*$ indicates complex conjugate and $\mathbf{p}\left(\mathbf{r}_{i}\right)=  \epsilon_{0}\alpha\mathbf{E}_\mathrm{loc}\left(\mathbf{r}_{i}\right)$ is the induced dipole moment of the $i$th atom. The optical theorem indicates that, for atoms without any nonradiative transitions, the extinction and scattering cross sections are identical, i.e., $C_{\mathrm{sca}}=C_{\mathrm{ext}}$~\cite{Milonni2008}. The total field at the position of the $i$th atom $\mathbf{E}_\mathrm{loc}\left(\mathbf{r}_{i}\right)$ is the sum of the incident field and the scattered field from the other atoms. The electric dipole at position $\mathbf{r}_{j}$ radiates an electromagnetic field which when measured at $\mathbf{r}_{i}$ can be calculated from $\mathbf{G}\left(\mathbf{r}_{i},\mathbf{r}_{j}\right)\mathbf{p}\left(\mathbf{r}_{j}\right)$, where $\mathbf{G}\left(\mathbf{r}_{i},\mathbf{r}_{j}\right)$ is Green's tensor~\cite{tai1994dyadic,Jackson1999}. Therefore, the dipole moment of the atom can be calculated from the following coupled equations ~\cite{SM,foldy1945,Mulholland:94}:

\begin{equation}
\mathbf{p}\left(\mathbf{r}_{i}\right)=  \epsilon_{0}\alpha\left(\mathbf{E}_{\mathrm{inc}}\left(\mathbf{r}_{i}\right)+\sum\limits_{j=1,j\neq i}^{N}\mathbf{G}\left(\mathbf{r}_{i},\mathbf{r}_{j}\right)\mathbf{p}\left(\mathbf{r}_{j}\right)\right). \label{eq:CDT}
\end{equation}
Equation~\eqref{eq:CDT} can be solved numerically for an arbitrary geometry of the atoms and the driving field. 

\textit{Superscattering.—} 
 In general, the maximum scattering cross section of an isotropic scatterer is given by $C_{\mathrm{sca},j}^{\mathrm{max}}=\left(2j+1\right)\frac{\lambda^{2}}{2\text{\ensuremath{\pi}}}$, where, $j$ is the order of the multipole; e.g., $j=1$, $2$, and $3$, for dipoles, quadrupoles, and octupoles, respectively~\cite{Ruan2010,Ruan2011}. For example, the maximum scattering cross section of a single dipolar scatterer (such as a two-level atom considered in this Letter) occurs at the resonance frequency and is limited to $3\lambda^2/2\pi$~\cite{foot2005,Zumofen:2008}. However, it is possible to engineer the structure of the scatterer to align the resonance frequencies of different multipoles, and thus, enhance the scattering cross section dramatically~\cite{Ruan2010,Ruan2011}. Here, we show that a particular arrangement of $N$ atoms can exhibit a scattering cross section significantly larger than $N\times\frac{3\lambda^{2}}{2\pi}$, although each individual atom supports only an electric dipole transition.
 
Consider $N$ atoms equally placed on the \textit{z}-axis and illuminated by an \textit{x}-polarized plane wave propagating in the $z$ direction as shown in the inset of Fig.~\ref{fig:Superscattering_N_atoms}. By using Eqs.~\eqref{eq:Ext_def}-\eqref{eq:CDT}, we plot the normalized maximum scattering cross section as a function of the distance between the atoms, $l$. At large atomic separations, i.e., $l>\lambda_{a}$, the atoms are nearly non-interacting and the maximum cross section is limited to $N\times\frac{3\lambda^{2}}{2\pi}$~(see the gray line in Fig.~\ref{fig:Superscattering_N_atoms}). However, the atomic antenna exhibits superscattering at one particular atomic spacing of $l\approx0.2\lambda_a$. Notably, the maximum scattering cross section for $N=5$ is approximately $15\times\frac{3\lambda^{2}}{2\pi}$. To achieve an even larger scattering cross section for a subwavelength atomic antenna, one can devise a three-dimensional atomic arrangement and obtain, for example,  $\ensuremath{C_{\mathrm{sca}}^{\mathrm{max}}\approx28\times\frac{3\lambda^{2}}{2\pi}}$ for an antenna composed of $N=16$ atoms (see Supplementary Material (SM) for more details~\cite{SM}).

\begin{figure}
    \centering
    \includegraphics[width=0.48\textwidth]{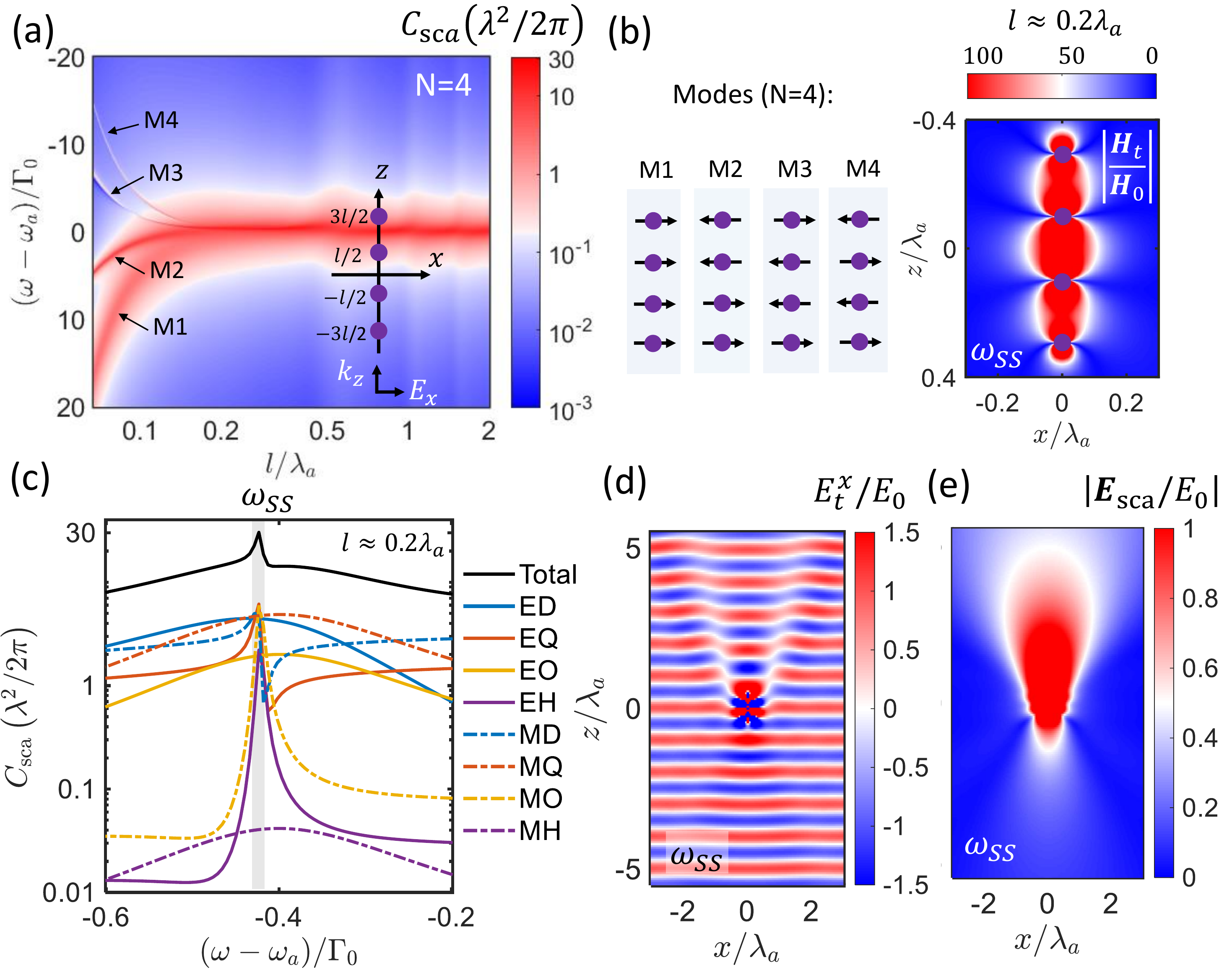}
    \caption{\textit{Superscattering in a tetramer:} (a) Scattering cross section normalized to $\text{\ensuremath{\lambda^{2}/2\pi}}$ and shown in logarithmic scale. Inset: Illustrates an atomic tetramer placed at $\mathbf{r}_{1,2}=\pm l/2\mathbf{e}_{z},\, \mathbf{r}_{3,4}=\pm3l/2\mathbf{e}_{z}$. (b) The tetramer exhibits four modes, $M_1$ to $M_4$. Total magnetic field distribution $\mathbf{H}_t$ at $\omega_{\rm SS}$, normalized to the incident magnetic field $H_0$. (c) Normalized scattering cross section of different electric and magnetic multipole moments as a function of frequency detuning for $l\approx0.2\lambda_a$. ED (MD), EQ (MQ) and EO (MO), EH (MH) indicate the electric (magnetic) dipole, quadrupole, octupole, and hexadecapole moments, respectively. (d)-(e) Real part of the normalized total electric field distribution $E_t^x$ (the $x$-component) and normalized scattered field distribution $\mathbf{\left|E_{{\rm sca}}\right|}$ at $\omega_{\mathrm{SS}}$, respectively.
    }
    \label{fig:Superscattering}
\end{figure}
In order to understand the nature of this enhancement in the scattering cross section, we examine the case of an atomic tetramer with $N=4$~[see the inset of Fig.~\ref{fig:Superscattering}(a)]. The normalized scattering cross section as a function of frequency detuning and distance $l$ is shown in Fig.~\ref{fig:Superscattering}(a). The tetramer exhibits four modes~[Fig.~\ref{fig:Superscattering}(b)]. M1 is a superradiant mode, where all dipoles oscillate in phase, and exhibits a large linewidth compared to the natural linewidth of an isolated atom~\cite{Dicke:1954,Lehmberg1970,Friedberg1973,Haroche1982,Asenjo2017}.
In contrast, M4 is subradiant with a very narrow linewidth.

Figure~\ref{fig:Superscattering}(c) shows the normalized scattering cross section as a function of frequency detuning for $l=0.2\lambda_a$ and the contribution of each multipole moment to the scattering cross section. Remarkably, the maximum scattering cross section of different multipole moments occur, approximately, at the superscattering frequency, $\omega_{\mathrm{SS}}$. The total scattering cross section is a superposition of the cross sections associated with effective electric and magnetic dipole, quadrupole and octupole moments. Therefore, the scattering cross section of the tetramer far exceeds the scattering cross section of a single atom. The strong magnetic field in Fig.~\ref{fig:Superscattering}(b) testifies to induced magnetic multipolar response at $\omega_{\rm SS}$. Using $\mathbf{E}_{\mathrm{sca}}\left(\mathbf{r}\right)=\sum\limits_{i=1}^{N}\mathbf{G}\left(\mathbf{r},\mathbf{r}_{i}\right)\mathbf{p}\left(\mathbf{r}_{i}\right)$, 
the scattered and the total (sum of incident and scattered) fields of the atomic antenna in free space can be calculated at $\mathbf r$. Figure~\ref{fig:Superscattering}(d)-(e) show the scattered and total field distribution at the superscattering frequency $\omega_{\mathrm{SS}}$. The scattering field is very large even at the far-field of the tetramer ($>>\lambda$) and the total field is strongly perturbed~[see Fig.~\ref{fig:Superscattering}(d)]. Therefore, a higher extinction can be achieved compared to that of a single atom~\cite{Zumofen:2008,Wrigge2008}. 
\begin{figure}
    \centering
    \includegraphics[width=0.48\textwidth]{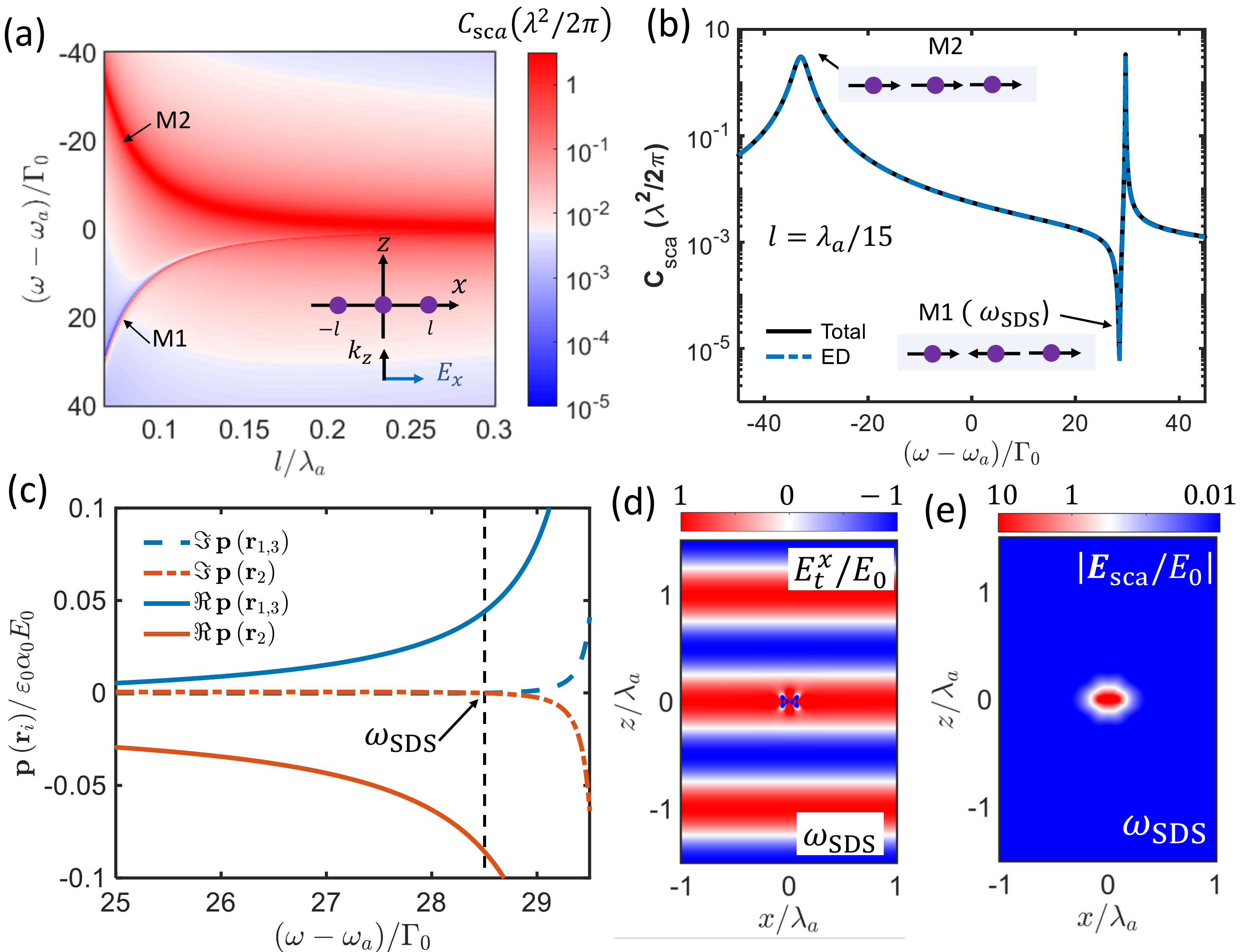}
    \caption{\textit{Scattering dark state as a nonradiating source:} (a) Scattering cross section (normalized to $\text{\ensuremath{\lambda^{2}/2\pi}}$ and shown in logarithmic scale) as a function of frequency detuning and the distance between the atoms $l$. Inset: schematic drawing of an atomic trimer composed of atoms at $\mathbf{r}_{1,3}=\mp l\mathbf{e}_{x}$, and $\mathbf{r}_{2}=0$, and the plane wave excitation.
    (b) Normalized scattering cross section as a function of frequency detuning for $l=\lambda_a/15$. The dashed line shows the contribution of the electric dipole moment. The trimer exhibits two modes $M_1$ and $M_2$ when illuminating by a plane wave. (c) Real and imaginary part of the induced dipole moment of each atom as a function of frequency detuning. At $\omega_{SDS}$, the individual atoms have nonzero dipole moments.
 (d) Real part of the normalized total field distribution $E_t^x$ at the frequency of scattering dark state,  $\omega_{\mathrm{SDS}}$. (e) Normalized scattered field distribution $\left|\mathbf{E}_{\mathrm{sca}}\right|$ at $\omega_{\mathrm{SDS}}$ (shown in logarithmic scale).}
    \label{fig:SDS}
\end{figure}


\textit{Scattering dark state (SDS).—} While each atom is excited close to the resonance and scatters the photons individually, it is possible, in some certain conditions, to achieve a negligible overall scattering from the structure. This phenomenon, known as the scattering dark state, has been investigated in nanoparticles~\cite{Devaney:1973,Hsu:2014}. In an SDS, the current distribution, and hence the electromagnetic field, is nonzero or even very large in the vicinity of the structure. However, similar to a nonradiating source~\cite{Devaney:1973}, the current distribution does not radiate to the far field. Therefore, such a structure becomes invisible under conditions of an SDS~\cite{Hsu:2014}. Here, we study scattering dark states in an atomic antenna with a nonzero induced current and very large near fields. Let us consider an atomic trimer placed on the \textit{x}-axis and illuminated by an \textit{x}-polarized plane wave propagating in the $z$ direction~[shown in the inset of Fig.~\ref{fig:SDS}(a)]. The scattering (extinction) cross section of the atomic trimer can be calculated using $C_{\mathrm{sca}}=k\mathrm{Im}\left(\alpha_{\mathrm{sum}}\right)$, where $\alpha_{\mathrm{sum}}$ read as~\cite{SM}
\begin{eqnarray}
\alpha_{\mathrm{sum}}& = &\sum\limits_{i=1}^{3}\alpha_{i}=\alpha\frac{\epsilon_{0}\alpha(\text{\ensuremath{\beta_{2}}}-4\beta_{1})-3}{\epsilon_{0}\alpha\left(2\epsilon_{0}\alpha\beta_{1}^{2}+\text{\text{\ensuremath{\beta_{2}}}}\right)-1},\label{eq:C_sca_SDS}
\end{eqnarray}
where $\alpha_i$ is the effective electric polarizability of each atom at its position. $\beta_1=G_{12}^{xx}\left(kl\right)$ and $\beta_2=G_{13}^{xx}\left(2kl\right)$ are components of Green's tensor~\cite{SM}.

The normalized scattering cross section as a function of frequency detuning and distance between the atoms $l$ is plotted in Fig.~\ref{fig:SDS}(a). The trimer can support multiple modes. However, only two modes can couple to the driving plane wave. These modes are subradiant and superradiant modes~\cite{Dicke:1954,Lehmberg1970,Friedberg1973,Haroche1982,Jenkins:2012,Feng2013,Bettles2015,Bettles2016Cooperative,Asenjo2017} and are labeled by M1 and M2 in Fig.~\ref{fig:SDS}, respectively. Figure~\ref{fig:SDS}(b) shows the normalized scattering cross section as a function of frequency detuning for $l=\lambda_a/15$. At the scattering dark state, $\omega_{\mathrm{SDS}}\approx28.5\Gamma_0$, the trimer is transparent and we observe that the scattering cross section is significantly suppressed by \textit{five} orders of magnitude compared to the on-resonance excitation. Therefore, the trimer becomes invisible to the incident field at the scattering dark state as the scattered field of the atoms interfere destructively in the far field. By using a multipole expansion of the induced current at $\mathbf{r}=0$, we calculate the contribution of electric dipole and higher-order multipoles to the scattering cross section~\cite{Alaee:2018}. The atomic trimer supports only an electric dipole moment, i.e., $p_{\mathrm{eff}}^{x}\approx\epsilon_{0}E_{0}\alpha_{\rm sum},$ [see the blue dashed line in Fig.~\ref{fig:SDS}(b), i.e., ED] and all the higher multipoles are negligible~\cite{SM}. Note that at $\omega_{\rm SDS}$ all the effective multipole moments vanish which is the condition for the SDS~\cite{Devaney:1973}. While the scattered field is nearly zero, the induced dipole of each atom and the induced current are not negligible as shown in Fig.~\ref{fig:SDS}(c). 

At the transparency frequency $\omega_{\mathrm{SDS}}$, we observe an extremely small perturbation to the incident plane wave~[Fig.~\ref{fig:SDS}(d)]. Remarkably, the field distribution at $\omega_{\mathrm{SDS}}$ implies that the electromagnetic fields generated inside the scatterer is nonzero; and indeed, it is one order of magnitude larger than the incident electric field~[Fig.~\ref{fig:SDS}(e)]. At $\omega_{\mathrm{SDS}}$, the induced current of the trimer is a nearly nonradiating source. In the supplementary material~\cite{SM}, we investigate antennas consisting of larger number of atoms (e.g. pentamer and heptamer). We show that for an antenna with $n$ excited modes there are $n-1$ scattering dark states. In other words, only one resonant mode radiates while the other resonances do not radiate because of vanishing induced multipole moments~\cite{Devaney:1973,Hsu:2014}. 
\begin{figure}
    \centering
    \includegraphics[width=0.48\textwidth]{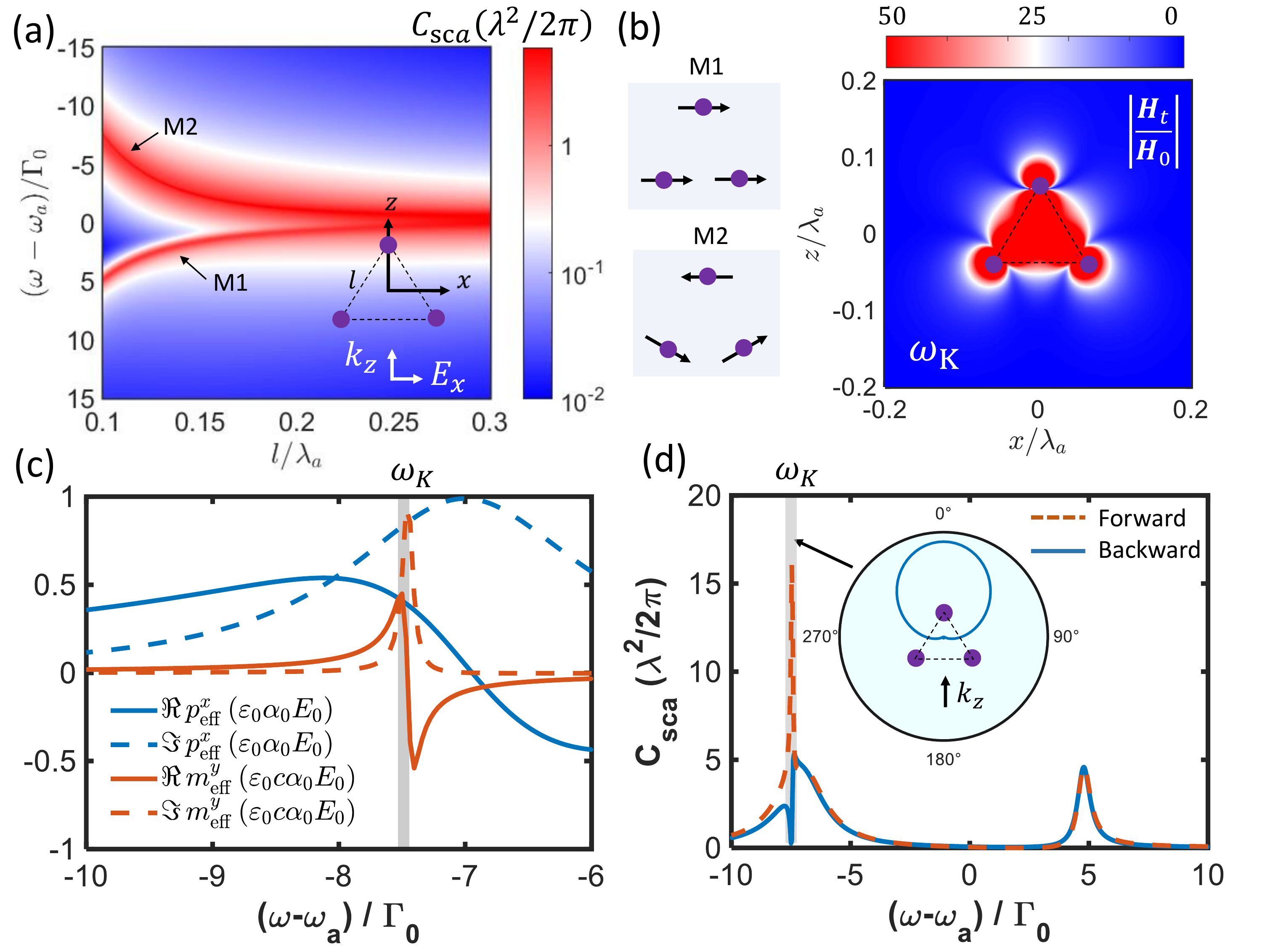}
    \caption{\textit{Kerker condition to achieve zero backscattering:} (a) Total scattering cross section normalized to $\text{\ensuremath{\lambda^{2}/2\pi}}$ as a function of the distance between the atoms. Inset: schematic drawing of an atomic trimer in an equilateral triangle composed of atoms at $\mathbf{r}_{1,2}=\pm l/2\mathbf{e}_{x}-\frac{\sqrt{3}}{6}l\mathbf{e}_{z},$ and $\mathbf{r}_{3}=\frac{\sqrt{3}}{3}l\mathbf{e}_{z}$. (b) Normalized total magnetic field distribution at $\omega_{\rm K}$. The trimer exhibits two modes $M_1$ and $M_2$. (c) The real and imaginary parts of the induced electric $p_{\rm eff}^{x}$ and magnetic $m_{\rm eff}^{y}$ dipole moments calculated from Eq.~\eqref{eq:Kerker_p_m}. The Kerker condition is met at $\omega_{\rm K}$ where the real and imaginary parts of the induced dipole moments satisfy the condition $p_{\rm eff}^{x}=m_{\rm eff}^{y}/c$, independently. (d) Normalized forward (red line) and backward (blue line) scattering cross section at $l = 0.1\lambda_{a}$ calculated from Eq.~\eqref{C_froward_backward}. Inset: the radiation pattern at the Kerker condition, which shows zero backward scattering.}
    \label{fig:Kerker}
\end{figure}

\textit{Kerker effect.—} An atom or a scatterer with only an electric dipole transition has an omnidirectional radiation pattern, i.e. scatters lights in both forward and backward directions. However, a Huygens’ scatterer, a scatterer with balanced electric and magnetic dipole moments, can exhibit a directional radiation pattern with zero backscattering. This effect is also known as the first Kerker condition~\cite{kerker1983}. The directional radiation with zero backscattering relies on constructive and destructive interference of the scattered field of the induced electric and magnetic dipole moments in the forward and backward directions, respectively. The Kerker condition has been studied in engineered high index dielectric and metallic nanoparticles~\cite{kerker1983,Alaee_kerker:15}. In the following, we show that an atomic trimer in an equilateral triangle configuration~[see the inset of Fig.~\ref{fig:Kerker}(a)] can satisfy the first Kerker condition, although the constituent atoms support only an electric dipole transition. The normalized scattering cross section as a function of frequency detuning and the distance between the atoms, $l$, is shown in Fig.~\ref{fig:Kerker}(a). The trimer exhibits two modes as shown with M1 and M2 in Fig.~\ref{fig:Kerker}(b)~\cite{SM,Feng2013}. We will show below that the M2 mode exhibits a strong magnetic response~[see the magnetic field enhancement in Fig.~\ref{fig:Kerker}(b)] and satisfies the Kerker condition. By decreasing the distance $l$, the separation between the two modes increases because of the strong near field coupling. 

To understand the underlying physics of the Kerker condition, we note that the proposed structure supports both \textit{effective} electric and magnetic dipole moments, which are given by~\cite{Alaee:2018, SM}
\begin{eqnarray}
p_{\mathrm{eff}}^{x} & = & \left(2p_{1}^{x}+p_{3}^{x}\right)j_{0}\left(u\right)+\frac{1}{4}\left(5p_{1}^{x}-2p_{3}^{x}\right)j_{2}\left(u\right),\nonumber \\
m_{\mathrm{eff}}^{y} & = & \frac{3i}{2}c\left[\left(p_{1}^{x}-p_{3}^{x}\right)j_{1}\left(u\right)+\sqrt{3}p_{2}^{z}j_{2}\left(u\right)\right],\label{eq:Kerker_p_m}
\end{eqnarray}
where $j_{n}\left(u\right)$ are spherical Bessel
functions, $u=\sqrt{3}kl/{3}$, and the superscripts indicate the direction of the dipole moments. $p_{1,3}^{x}$ and $p_{2}^{z}$ are the electric dipole moments of the individual atoms, along the $x$- and $z$-axis, and are calculated from Eq.~\eqref{eq:CDT}. Note that in the derivation of Eq.~\eqref{eq:Kerker_p_m}, we used $p_{1}^{x}=p_{2}^{x}$ and $p_{1}^{z}=-p_{2}^{z}$ due to the symmetry of the trimer~\cite{SM}.
Then, the forward and backward scattering cross sections for this atomic antenna can be calculated from~\cite{SM}
\begin{equation}
C_{\mathrm{sca}} = k^{4}\left|p_{\mathrm{eff}}^{x}\pm m_{\mathrm{eff}}^{y}/c\right|^{2}/\left(4\pi\epsilon_{0}^{2}\left|\mathbf{E}_{\mathrm{inc}}\right|^{2}\right),\label{C_froward_backward}
\end{equation}
where, $+$ and $-$ are used for forward and backward cross sections, respectively. Therefore, Eq.~\eqref{C_froward_backward} shows clearly that the first Kerker condition is fulfilled when $p_{\rm eff}^{x}=m_{\rm eff}^{y}/c$. Figure~\ref{fig:Kerker}(c) plots the real and imaginary parts of the effective electric and magnetic dipole moments. It can be seen that at the Kerker frequency, $\omega_{\rm K}$, both the real and imaginary parts of the induced dipole moments satisfy the Kerker condition independently. Hence, at $\omega_{\rm{K}}$, the forward scattering is significantly enhanced while the backward scattering is considerably suppressed~[see Fig.~\ref{fig:Kerker}(d)].
 

A subwavelength ensemble of \textit{randomly} distributed atoms scatters less than a single atom close to resonance~\cite{Schilder:2020}. However, we showed that by arranging the atoms in particular geometries, one can control scattering to achieve extremely small or large scattering cross sections, the so called scattering dark state and superscattering, respectively. Moreover, we demonstrated that a particular geometry of the atoms fulfills the Kerker condition and scatters light in the forward direction. We showed that these geometries exhibit higher-order electric and magnetic multipole moments, although the individual atoms support only electric dipole transitions. We have employed these induced multipole moments to control the scattering. The treatment used in our work is an excellent approximation to the full quantum model in the weak excitation regime. Therefore, our considerations can also be realised in the quantum regime and pave the way towards the generation and manipulation of photonic quantum states, e.g., realization of single photon emitters~\cite{Chou2004,Black2005} and quantum memory~\cite{Duan2001,Reimann2015}.


\textbf{Acknowledgments.—}This work was partially supported by the Max Planck Society. R. A. is grateful to Boris Braverman for helpful discussions and acknowledges the support of the Alexander von Humboldt Foundation through the Feodor Lynen Fellowship. R.A., A.S., and R.W.B. acknowledge support through the Natural Sciences and Engineering Research Council of Canada, the Canada Research Chairs program, and the Canada First Research Excellence Fund.



\appendix
\begin{widetext}

\section{Atomic polarizability and coupled-dipole equations}
Let us consider an atomic antenna composed of natural atoms with only
$\textit{electric}$ dipole transition moments and illuminated by a plane wave~[see Fig.~\ref{fig:AtomicAntennas}]. The atoms are arranged at small distances such that the entire size of the antenna is smaller than the wavelength of the light,
i.e., $D<\lambda$. We consider
the weak-excitation limit where the atomic response is isotropic and
linear. The electric polarizability of each atom amounts to $\alpha\left(\omega\right)=-\left(\alpha_{0}\Gamma_{0}/2\right)/\left[\delta+i\left(\Gamma_{0}+\Gamma_{\mathrm{nr}}\right)/2\right]$,
where $\Gamma_{0}$ is the radiative linewidth of the atomic transition at frequency $\omega_{a}$, and $\delta_a=\omega-\omega_{a}\ll\omega_{a}$
represents the frequency detuning between the illumination and the
atom, $\alpha_{0}=6\pi/k^{3}$ and $k$ is the wavenumber~\cite{lagendijk1996,Lagendijk1998}.
We assume elastic scattering events and therefore the non-radiative
decay rate is zero, i.e., $\Gamma_{\mathrm{nr}}=0$. The induced displacement
volume current density for the atomic antenna can be written as
\begin{equation}
\mathbf{J}\left(\mathbf{r},\omega\right)  =  -i\omega\sum_{i=1}^{N}\mathbf{p}\left(\mathbf{r}_{i}\right)\delta\left(\mathbf{r}-\mathbf{r}_{i}\right),\label{eq:Current_atoms}
\end{equation}
where $\delta$ is the Dirac delta function and $\mathbf{p}\left(\mathbf{r}_{i}\right)$
is the induced electric dipole moment of the $i$th atom placed at $\mathbf{r}=\mathbf{r}_{i}$~[see Fig.~\ref{fig:AtomicAntennas}]. In Eq.~(\ref{eq:Current_atoms}),
we assumed $e^{-i\omega t}$ as a time harmonic variation. The induced
dipole moment of the $i$th atom can be obtained by using the coupled-dipole
equations~\cite{lagendijk1996,Lagendijk1998,Alaee:2017Review}
\begin{eqnarray}
\mathbf{p}\left(\mathbf{r}_{i}\right) & = & \epsilon_{0}\alpha_{i}\left[\mathbf{E}_{\mathrm{inc}}\left(\mathbf{r}_{i}\right)+\underset{i\neq j}{\sum}\mathbf{G}\left(\mathbf{r}_{i},\mathbf{r}_{j}\right)\mathbf{p}\left(\mathbf{r}_{j}\right)\right],\label{eq:CDT_S}
\end{eqnarray}
where $\mathbf{E}_{\mathrm{inc}}\left(\mathbf{r}_{i}\right)$ is the
incident field at the atom position $\mathbf{r}_i$, $\alpha_{i}$ is the atomic polarizability, and $\underset{i\neq j}{\sum}\mathbf{G}\left(\mathbf{r}_{i},\mathbf{r}_{j}\right)\mathbf{p}\left(\mathbf{r}_{j}\right)$
is the interaction field at $\mathbf{r}=\mathbf{r}_{i}$ created by all the other atoms. Green's function in 
Eq.~(\ref{eq:CDT_S}) reads as~\cite{tai1994dyadic,Jackson1999}
\begin{equation}
\text{\ensuremath{\mathbf{G}\left(\mathbf{r}_{i},\mathbf{r}_{j}\right)}=\ensuremath{\frac{3}{2\alpha_{0}\epsilon_{0}}e^{i\zeta}\left[\left(\frac{1}{\zeta}-\frac{1}{\zeta^{3}}+\frac{i}{\zeta^{2}}\right)\bar{\bar{{\bf I}}}+\left(-\frac{1}{\zeta}+\frac{3}{\zeta^{3}}-\frac{3i}{\zeta^{2}}\right)\mathbf{\mathbf{n}n}\right]}},
\end{equation}
where $\bar{\bar{{\bf I}}}$ is the identity dyadic, $\mathbf{n}=\frac{\mathbf{r}_{i}-\mathbf{r}_{j}}{\left|\mathbf{r}_{i}-\mathbf{r}_{j}\right|}$,
and $\zeta=k\left|\left(\mathbf{r}_{i}-\mathbf{r}_{j}\right)\right|$. The Cartesian components of Green's tensor are given by
\begin{eqnarray}
G^{\mu\nu}\left(\zeta=k\left|\mathbf{r}_{i}-\mathbf{r}_{j}\right|\right) & = & \frac{3}{2\alpha_{0}\epsilon_{0}}e^{i\zeta}\left[g_{1}\left(\zeta\right)\delta_{\mu\nu}+g_{2}\left(\zeta\right)\frac{\zeta_{\mu}\zeta_{\nu}}{\zeta^{2}}\right],\nonumber \\
g_{1}\left(\zeta\right) & = & \left(\frac{1}{\zeta}-\frac{1}{\zeta^{3}}+\frac{i}{\zeta^{2}}\right),\,\,\,\,\,\,\,\,\,
g_{2}\left(\zeta\right)  =  \left(-\frac{1}{\zeta}+\frac{3}{\zeta^{3}}-\frac{3i}{\zeta^{2}}\right),\label{eq:GreenFunction}
\end{eqnarray}
where $\mu,\ \nu$  $\in {x, y, z}$. 

For atoms with zero non-radiative decay $\Gamma_{\mathrm{nr}}=0$,
we get $\mathrm{Im}\left[1/\alpha\right]=-1/\alpha_{0}$. Using conservation
of energy, one can show that the scattered power is equal to the extracted
power $P_{\mathrm{sca}}=P_{\mathrm{ext}}$. Thus, the scattering cross
section oand the extinction cross section of the atoms are identical and can be calculated from~\cite{Mulholland:94,Lagendijk1998}
\begin{eqnarray}
C_{\mathrm{sca}} & = & C_{\mathrm{ext}}=\frac{k}{\epsilon_{0}\left|E_{0}\right|^{2}}\sum_{i=1}^{N}\mathrm{Im}\left[\mathbf{p}\left(\mathbf{r}_{i}\right)\cdot\mathbf{E}_{\mathrm{inc}}^{*}\left(\mathbf{r}_{i}\right)\right].\label{eq:C_sca}
\end{eqnarray}

\begin{figure}
\begin{centering}
\includegraphics[width=5cm]{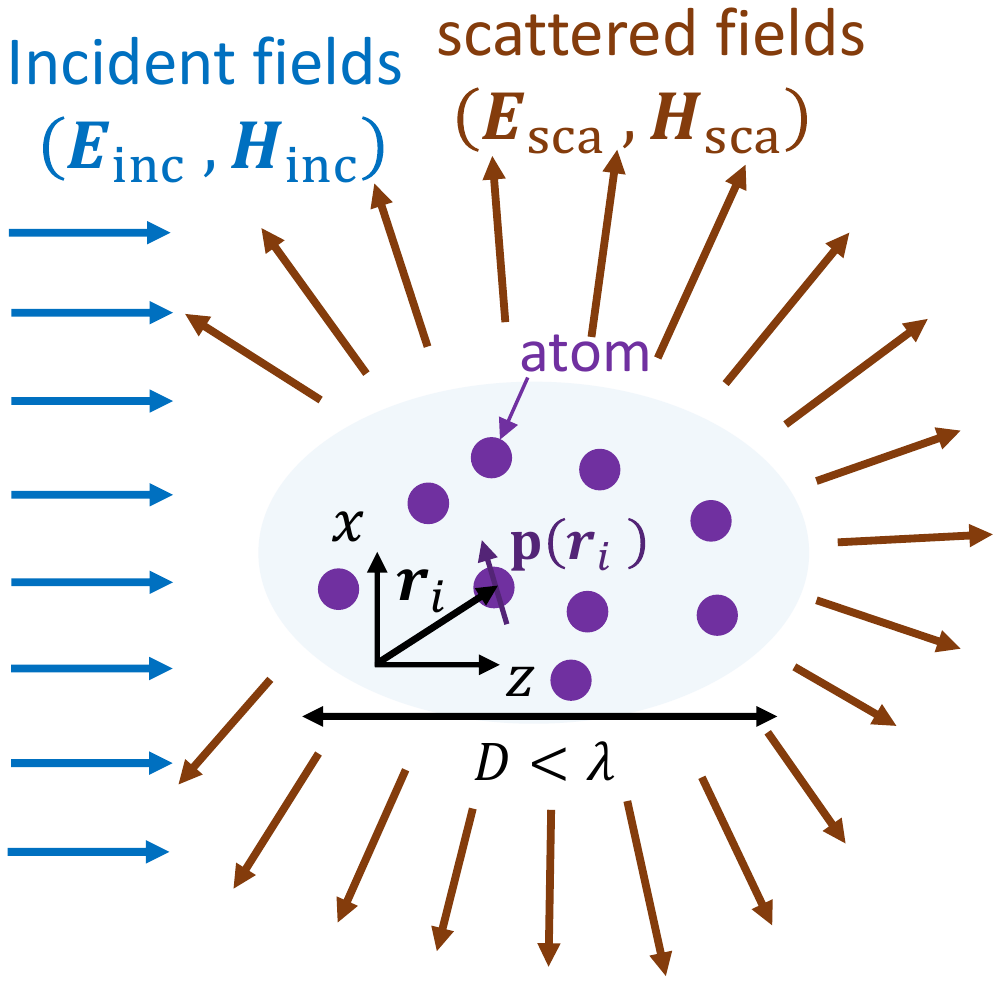}
\par\end{centering}
\caption{\textit{Subwavelength atomic antenna:} sketch of a quantum antenna
composed of natural atoms with only $\textit{electric}$ dipole transition
moments arranged at distances smaller than the wavelength of light,
i.e., $D<\lambda$, $D$ is the overall size of the antenna.\label{fig:AtomicAntennas}}
\end{figure}

\begin{figure}
\begin{centering}
\includegraphics[width=10cm]{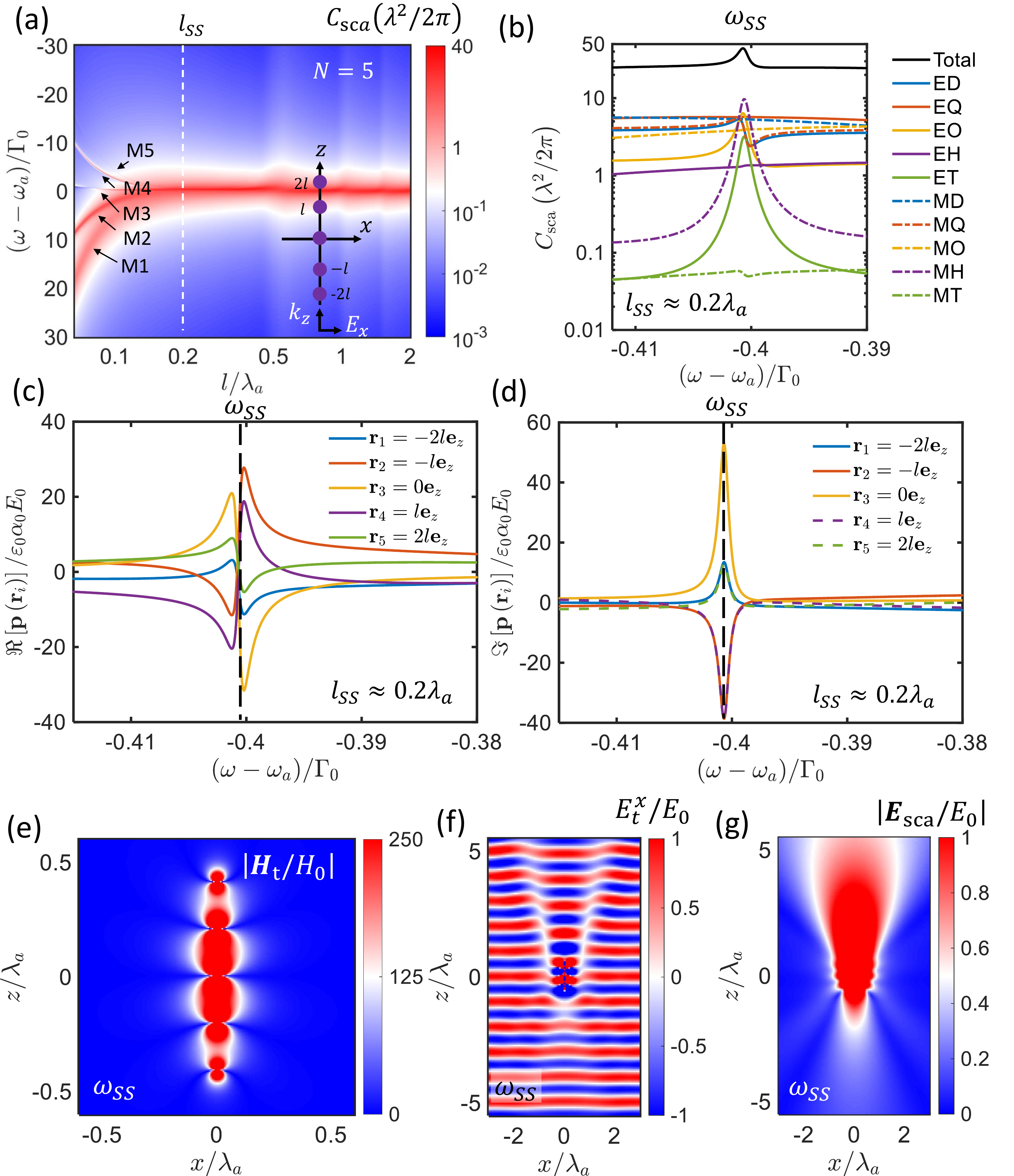}
\par\end{centering}
\caption{\textit{Superscattering in a pentamer:} (a) Scattering cross section normalized to $\text{\ensuremath{\lambda^{2}/2\pi}}$ and shown in logarithmic scale. The inset illustrates the atoms placed at $\mathbf{r}_{1,5}=\mp 2l\mathbf{e}_{z},\, \mathbf{r}_{2,4}=\mp l\mathbf{e}_{z},\ \mathbf{r}_{3}= 0\mathbf{e}_{z}$. The pentamer exhibits five modes, $M_1$ to $M_5$. (b) Normalized scattering cross section of different electric and magnetic multipole moments as a function of frequency detuning for $l_{\rm SS}\approx0.2\lambda_a$. ED (MD), EQ (MQ), EO (MO), EH (MH), and ET (MT) indicate the electric (magnetic) dipole, quadrupole, octupole, hexadecapole, and triakontadipole moments, respectively. (c)-(d) Real and imaginary part of the induced dipole moment of each atom as a function of frequency detuning at $l_{SS}$. (e) Total magnetic field distribution $\mathbf{H}_t$ at $\omega_{\rm SS}$, normalized to the incident magnetic field $H_0$. (f)-(g) Real part of the normalized total electric field distribution $E_t^x$ (the $x$-component) and normalized scattered field distribution $\mathbf{\left|E_{{\rm sca}}\right|}$ at $\omega_{\mathrm{SS}}$, respectively.}\label{fig:Superscattering_N5} 
\end{figure}

\begin{figure}
\begin{centering}
\includegraphics[width=10cm]{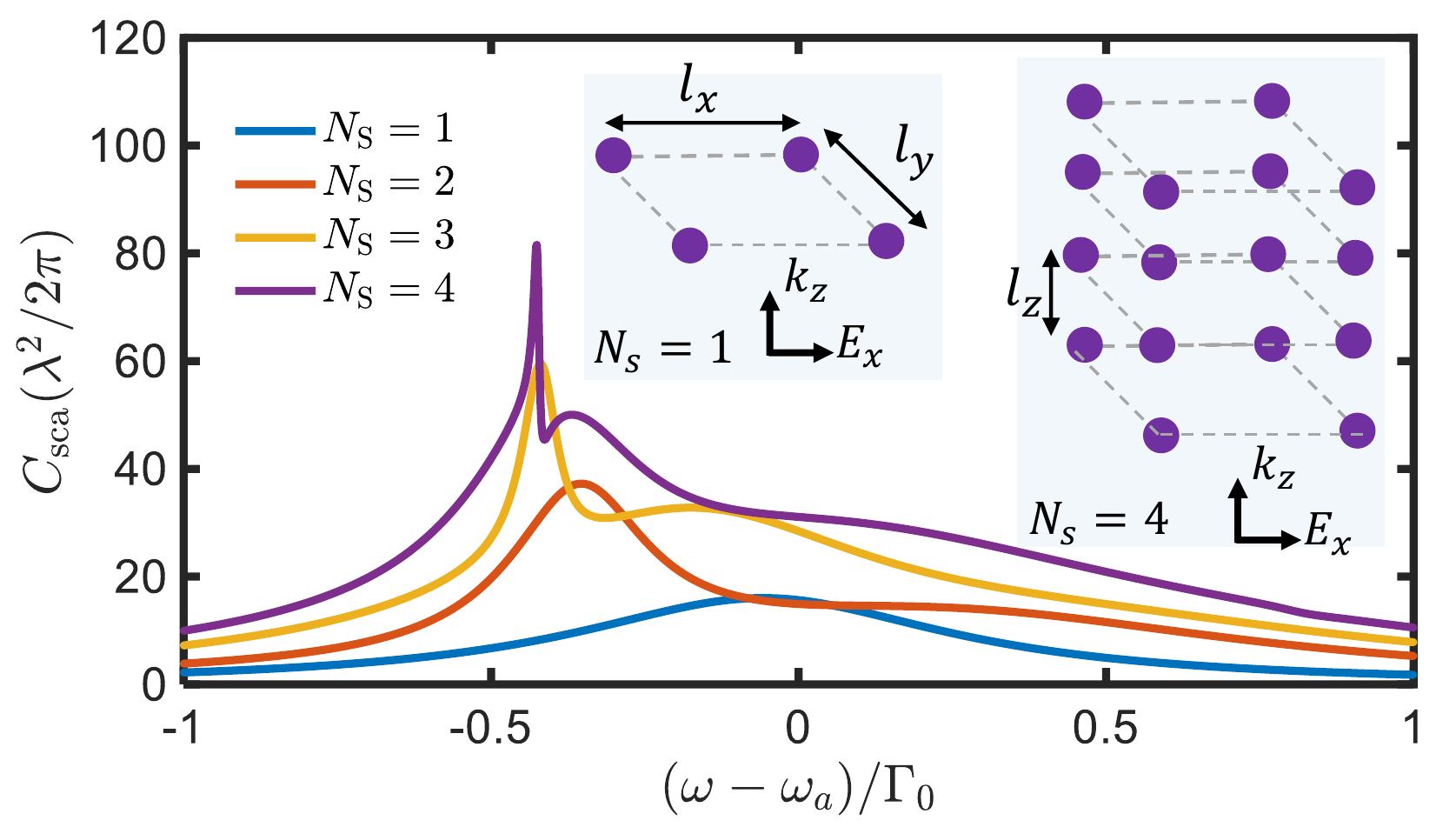}
\par\end{centering}
\caption{\textit{Superscattering in a three-dimensional subwavelength antenna:}
Scattering cross section normalized to $\text{\ensuremath{\lambda^{2}/2\pi}}$
as a function of frequency detuning~$\omega-\omega_{a}$ for
different layers of $2\times2$ atoms, i.e., $N_{s}=1,2,3,4$ (see the inset), where
$l_{x}=l_{y}=0.8\lambda_{a}$ and $l_{z}=0.2\lambda_{a}.$ Note that
the size of the antenna is smaller than the wavelength~($l_{x},l_{y},(N_s-1) \times l_{z}<\lambda$). Inset illustrates the three-dimensional subwavelength antenna for $N_{s}=1$, and $N=4$.
\label{fig:SS_3D}}
\end{figure}

\section{Superscattering}

In this section, we study supperscattering in one-dimensional and three-dimensional subwavelength atomic antennas. 
\subsection{One-dimensional subwavelength antenna}
Let us consider an atomic pentamer with $N=5$~[see the inset of Fig.~\ref{fig:Superscattering_N5}(a)]. The normalized scattering cross section as a function of frequency detuning $\omega-\omega_a$ and distance $l$ is shown in Fig.~\ref{fig:Superscattering_N5}(a). The pentamer exhibits five modes~[see the inset of Fig.~\ref{fig:Superscattering_N5}(a)]. M1 is a superradiant mode, where all dipoles oscillate in phase, and exhibit a large linewidth compared to the natural linewidth of an isolated atom.
In contrast, M5 is subradiant with a very narrow linewidth. Figure~\ref{fig:Superscattering_N5}(b) shows the normalized scattering cross section as a function of frequency detuning for $l\approx0.2\lambda_a$ and the contribution of each multipole moment to the scattering cross section. The maximum scattering cross section of different multipole moments occur, approximately, at the superscattering frequency, $\omega_{\mathrm{SS}}$. The total scattering cross section is a superposition of the cross sections associated with effective multipole moments. Therefore, the scattering cross section of the pentamer far exceeds the scattering cross section of a single atom. Figure~\ref{fig:Superscattering_N5} (c)-(d) depicts the induced dipole of each atom. At $\omega_{\rm SS}$, the induced dipole moment of each atom is at resonance. Each dipole is out-of phase with respect to neighbouring atoms which explains the strong magnetic field in Fig.~\ref{fig:Superscattering_N5}(e). This strong magnetic field testifies that the pentamer exhibits induced \textit{magnetic} multipolar response at $\omega_{\rm SS}$. Figure~\ref{fig:Superscattering_N5}(f)-(g) shows the scattered and total (sum of incident and scattered) field distribution at the superscattering frequency $\omega_{\mathrm{SS}}$. The scattering field is very large even at the far-field of the pentamer and the total field is strongly perturbed. Therefore, a higher extinction can be achieved compared to that of a single atom. 
\subsection{Three-dimensional subwavelength antenna}
In the main text, we discussed superscattering from a one dimensional array of atoms. In this subsection, we consider superscattering from higher dimensional subwavelength atomic structures and show that even larger scattering cross sections can be achieved compared to the one-dimensional antenna. Figure~\ref{fig:SS_3D}
shows scattering cross section normalized to $\text{\ensuremath{\lambda^{2}/2\pi}}$
as a function of frequency detuning~$\omega-\omega_{a}$ for different
layers of $2\times2$ atoms. $N_{s}$ is the number of layers. It can be seen that the
scattering cross section of the 3D structure is significantly larger than of a single atom,
e.g., $\ensuremath{C_{\mathrm{sca}}^{\mathrm{max}}\approx28\times\frac{3\lambda^{2}}{2\pi}}$
for $N_{s}=4$ although the 3D antenna is composed of $N=16$ atoms~[See the inset of Fig.~\ref{fig:SS_3D}]. Figure~\ref{fig:SS_3D_Ns3} (b) depicts the normalized scattering cross section of the 3D antenna~[see Fig.~\ref{fig:SS_3D_Ns3} (a) for an antenna composed of $N=12$ atoms, i.e., $N_s=3$] as a function of frequency detuning and the contribution of each multipole moment to the scattering cross section. The total scattering cross section is a superposition of the cross sections associated with effective multipole moments. It can be seen that at $\omega_{SS}$ the antenna exhibits strong electric and magnetic response which leads to a very large scattering cross section $\ensuremath{C_{\mathrm{sca}}^{\mathrm{max}}\approx20\times\frac{3\lambda^{2}}{2\pi}}$. Figure~\ref{fig:SS_3D_Ns3}(c) shows the scattering and total (sum of incident and scattered, $x$-component) field distribution at the superscattering frequency $\omega_{\mathrm{SS}}$. The scattering field is very large even at the far-field and the total field is strongly perturbed.

\begin{figure}
\begin{centering}
\includegraphics[width=9cm]{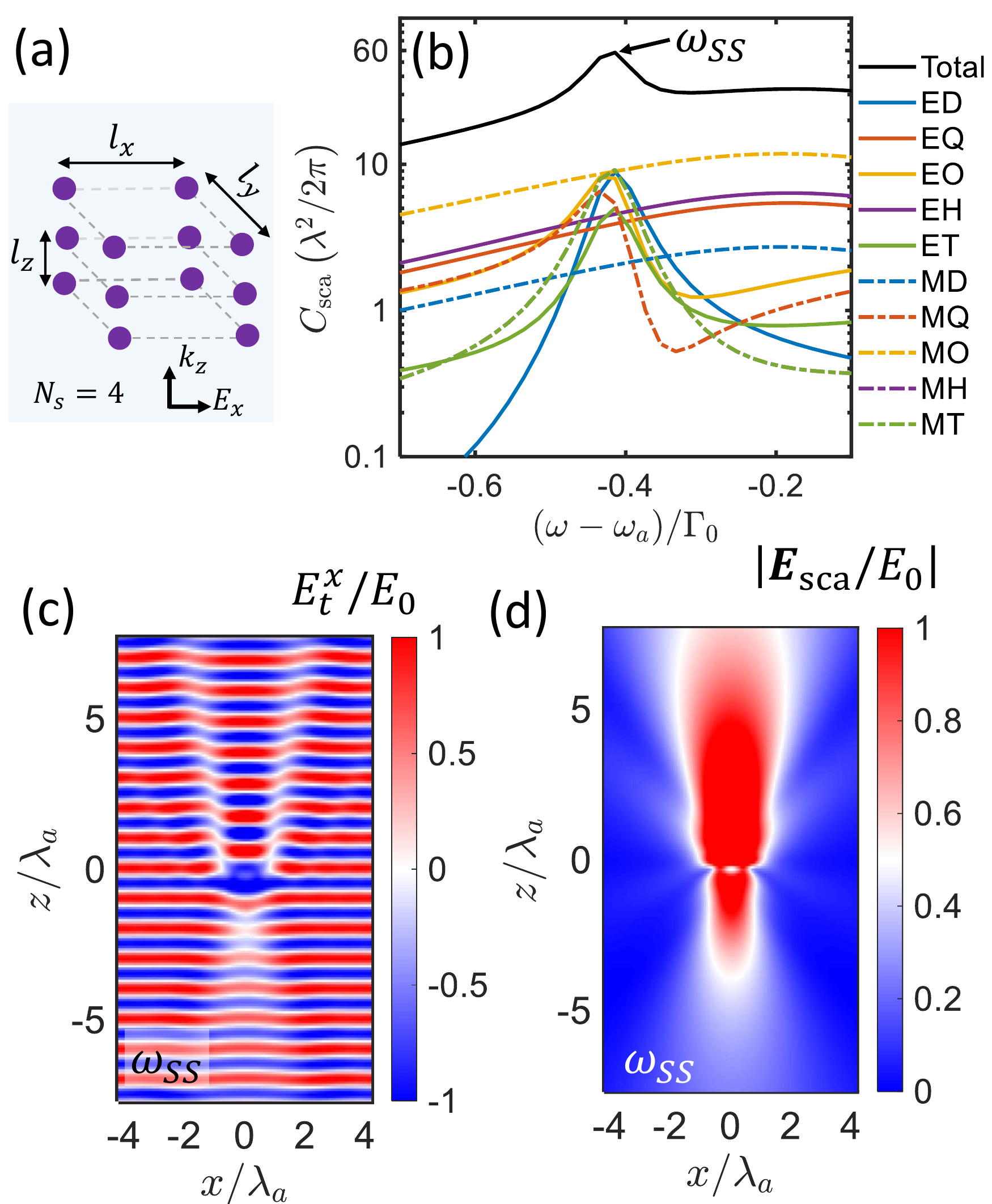}
\par\end{centering}
\caption{\textit{Superscattering in a three-dimensional subwavelength antenna, $N_s=3$:} (a) the three-dimensional subwavelength antenna composed of $N=$12 atoms ($N_{s}=3$), where
$l_{x}=l_{y}=0.8\lambda_{a}$ and $l_{z}=0.2\lambda_{a}$. (b) Normalized scattering cross section of different electric and magnetic multipole moments as a function of frequency detuning. ED (MD), EQ (MQ), EO (MO), EH (MH), and ET (MT) indicate the electric (magnetic) dipole, quadrupole, octupole, hexadecapole, and triakontadipole moments, respectively. (c)-(d) Real part of the normalized total electric field distribution $E_t^x$ (the $x$-component) and normalized scattered field distribution $\mathbf{\left|E_{{\rm sca}}\right|}$ at $\omega_{\mathrm{SS}}$, respectively.
\label{fig:SS_3D_Ns3}}
\end{figure}

\section{Scattering dark states (SDS)}
\subsection{Atomic trimer}
In this section, we derive effective multipole moments of an atomic trimer and its scattering cross section. The atomic trimer is placed on the $x$-axis with atoms at $\mathbf{r}_{1,3}=\pm l\mathbf{e}_{x}$,
$\mathbf{r}_{2}=0$ and is illuminated by an $x$-polarized plane wave
propagating in the $z$ direction [see Fig.~\ref{fig:SDS_Trimer}
(b)]. The induced displacement volume current density for the atomic
trimer can be written as
\begin{eqnarray}
\mathbf{J}\left(\mathbf{r},\omega\right) & = & -i\omega\left[p_{1}^{x}\delta\left(\mathbf{r}-\mathbf{r}_{1}\right)+p_{2}^{x}\delta\left(\mathbf{r}-\mathbf{r}_{2}\right)+p_{3}^{x}\delta\left(\mathbf{r}-\mathbf{r}_{3}\right)\right]\mathbf{e}_{x},\label{eq:J_SDS_trimer}
\end{eqnarray}
where $p_{1}^{x}=p_{3}^{x}$ due to symmetry~[see Fig.~\ref{fig:SDS_Trimer}]. The induced electric
dipole moment of each atom can be calculated using coupled equation, Eq.~(\ref{eq:CDT_S}) and read as
\begin{eqnarray}
p_{1}^{x} & = & p_{3}^{x}=-\epsilon_{0}\alpha\frac{\epsilon_{0}\alpha\beta_{1}+1}{\epsilon_{0}\alpha\left(2\epsilon_{0}\alpha\beta_{1}^{2}+\beta_{2}\right)-1}E_{0},\nonumber \\
p_{2}^{x} & = & \epsilon_{0}\alpha\frac{\epsilon_{0}\alpha\left[\beta_{2}-2\beta_{1}\right]-1}{\epsilon_{0}\alpha\left(2\epsilon_{0}\alpha\beta_{1}^{2}+\beta_{2}\right)-1}E_{0},\label{eq:P_SDS}
\end{eqnarray}
where $\beta_{1}$ and $\beta_{2}$ are defined by
\begin{eqnarray}
\beta_{1} & \overset{\varDelta}{=} & G_{12}^{xx}\left(\zeta=kl\right)=\frac{3}{2\alpha_{0}\epsilon_{0}}e^{ikl}\left(\frac{1}{kl}-\frac{1}{\left(kl\right)^{3}}+\frac{i}{\left(kl\right)^{2}}\right),\nonumber \\
\beta_{2} & \overset{\varDelta}{=} & G_{13}^{xx}\left(\zeta=2kl\right)=\frac{3}{2\alpha_{0}\epsilon_{0}}e^{i2kl}\left(\frac{1}{2kl}-\frac{1}{\left(2kl\right)^{3}}+\frac{i}{\left(2kl\right)^{2}}\right).
\end{eqnarray}
Using Eqs.~(\ref{eq:C_sca}) and ~(\ref{eq:P_SDS}), we obtain the
scattering cross section of the trimer~[i.e., Eq.~(3) of the main text]:
\begin{eqnarray}
C_{\mathrm{sca}} & = & \frac{k}{\epsilon_{0}\left|E_{0}\right|^{2}}\mathrm{Im}\left[\sum_{i=1}^{3}p_{i}E_{\mathrm{inc}}^{*}\left(\mathbf{r}_{i}\right)\right]=k\mathrm{Im}\left[\alpha\frac{\epsilon_{0}\alpha(\text{\ensuremath{\beta_{2}}}-4\beta_{1})-3}{\epsilon_{0}\alpha\left(2\epsilon_{0}\alpha\beta_{1}^{2}+\text{\text{\ensuremath{\beta_{2}}}}\right)-1}\right].\label{eq:C_ext_SDS_exact}
\end{eqnarray}

\textit{Multipole expansion at the center of the atomic trimer-.
}One can consider the trimer as a single antenna, effectively. The
induced \textit{effective} electric dipole moment of the trimer can
be obtained using the multipole expansion at the center of the trimer,
i.e., $\mathbf{r}=0$ and reads as~\cite{Alaee:2018,Alaee2019}
\begin{eqnarray}
p_{\mathrm{eff}}^{\beta} & = & -\frac{1}{i\omega}\left\{ \int dvJ_{\beta}j_{0}\left(kr\right)+\frac{k^{2}}{2}\int dv\left[3\left(\mathbf{r}\cdot\mathbf{J}\right)r_{\beta}-r^{2}J_{\beta}\right]\frac{j_{2}\left(kr\right)}{\left(kr\right)^{2}}\right\} ,\label{eq:p_ME-1}
\end{eqnarray}
where $\mathbf{J}$ is the induced displacement volume current in Eq.~\ref{eq:J_SDS_trimer} and $dv$ is the volume integral  $dv=dxdydz$ in Cartesian coordinates. $\beta=x,y,z$ and $j_{n}\left(kr\right)$ are spherical Bessel
functions. $\mathbf{p}_{\mathrm{eff}}=p_{\mathrm{eff}}^{x}\mathbf{e}_{x}$.
Now by substituting Eq.~(\ref{eq:J_SDS_trimer}) into Eq.~(\ref{eq:p_ME-1}),
we get
\begin{eqnarray}
p_{\mathrm{eff}}^{x} & = & -\frac{1}{i\omega}\left\{ \int J_{x}j_{0}\left(kr\right)dv+\frac{k^{2}}{2}\int dv\left[3\left(\mathbf{r}\cdot\mathbf{J}\right)x-r^{2}J_{x}\right]\frac{j_{2}\left(kr\right)}{\left(kr\right)^{2}}\right\} ,\nonumber \\
 & = & -\frac{1}{i\omega}\int J_{x}\left[j_{0}\left(kr\right)+\frac{1}{2}\left(3\frac{x^{2}}{r^{2}}-1\right)j_{2}\left(kr\right)\right]dv=\left[p_{1}^{x}+p_{3}^{x}\right]\left[j_{0}\left(kl\right)+j_{2}\left(kl\right)\right]+p_{2}^{x},\nonumber \\
 & = & \epsilon_{0}\alpha E_{0}\frac{\text{\ensuremath{\epsilon_{0}\alpha\beta_{2}}}-2\epsilon_{0}\alpha\beta_{1}-1-2(\epsilon_{0}\alpha\text{\text{\ensuremath{\beta_{1}}}}+1)\left[j_{0}\left(kl\right)+j_{2}\left(kl\right)\right]}{\epsilon_{0}\alpha\left[2\epsilon_{0}\alpha\beta_{1}^{2}+\beta_{2}\right]-1}.
\end{eqnarray}
Note that all the higher order multipoles are negligible for $kl\ll l$. Using Eq.~(\ref{eq:C_sca}), we calculate the scattering cross section
of the trimer
\begin{eqnarray}
C_{\mathrm{ext}} & \approx & \frac{k}{\epsilon_{0}\left|E_{0}\right|^{2}}\mathrm{Im}\left[p_{\mathrm{eff}}^{x}E_{\mathrm{inc}}^{*}\left(\mathbf{r}=0\right)\right],\nonumber \\
 & = & k\mathrm{Im}\left\{ \frac{\text{\ensuremath{\epsilon_{0}\alpha\beta_{2}}}-2\epsilon_{0}\alpha\beta_{1}-1-2(\epsilon_{0}\alpha\text{\text{\ensuremath{\beta_{1}}}}+1)\left[j_{0}\left(kl\right)+j_{2}\left(kl\right)\right]}{\epsilon_{0}\alpha\left[2\epsilon_{0}\alpha\beta_{1}^{2}+\beta_{2}\right]-1}\right\} ,\nonumber \\
 & \approx & k\mathrm{Im}\left[\alpha\frac{\epsilon_{0}\alpha(\text{\ensuremath{\beta_{2}}}-4\beta_{1})-3}{\epsilon_{0}\alpha\left(2\epsilon_{0}\alpha\beta_{1}^{2}+\text{\text{\ensuremath{\beta_{2}}}}\right)-1}\right].\label{eq:C_ext_SDS_app}
\end{eqnarray}
Clearly, Eq.~\ref{eq:C_ext_SDS_app} is identical to Eq.~\ref{eq:C_ext_SDS_exact}
and one can conclude that the trimer exhibits only an effective electric
dipole moment for $kl\ll1$.
In Eq.~(\ref{eq:C_ext_SDS_app}), we used the long-wavelength approximation,
i.e., $kl\ll1$, where the spherical Bessel functions can be approximated
by $j_{0}\left(kl\right)\approx1$ and $j_{2}\left(kl\right)\approx0$. Thus, the induced electric dipole moment can be written as
\begin{eqnarray}
p_{\mathrm{eff}}^{x} & \approx & \epsilon_{0}\alpha\frac{\epsilon_{0}\alpha\left(\text{\ensuremath{\beta_{2}}}-4\beta_{1}\right)-3}{\epsilon_{0}\alpha\left[2\epsilon_{0}\alpha\beta_{1}^{2}+\beta_{2}\right]-1}E_{0}.\label{eq:p_eff_SDS}
\end{eqnarray}

\begin{figure}
\begin{centering}
\includegraphics[width=5cm]{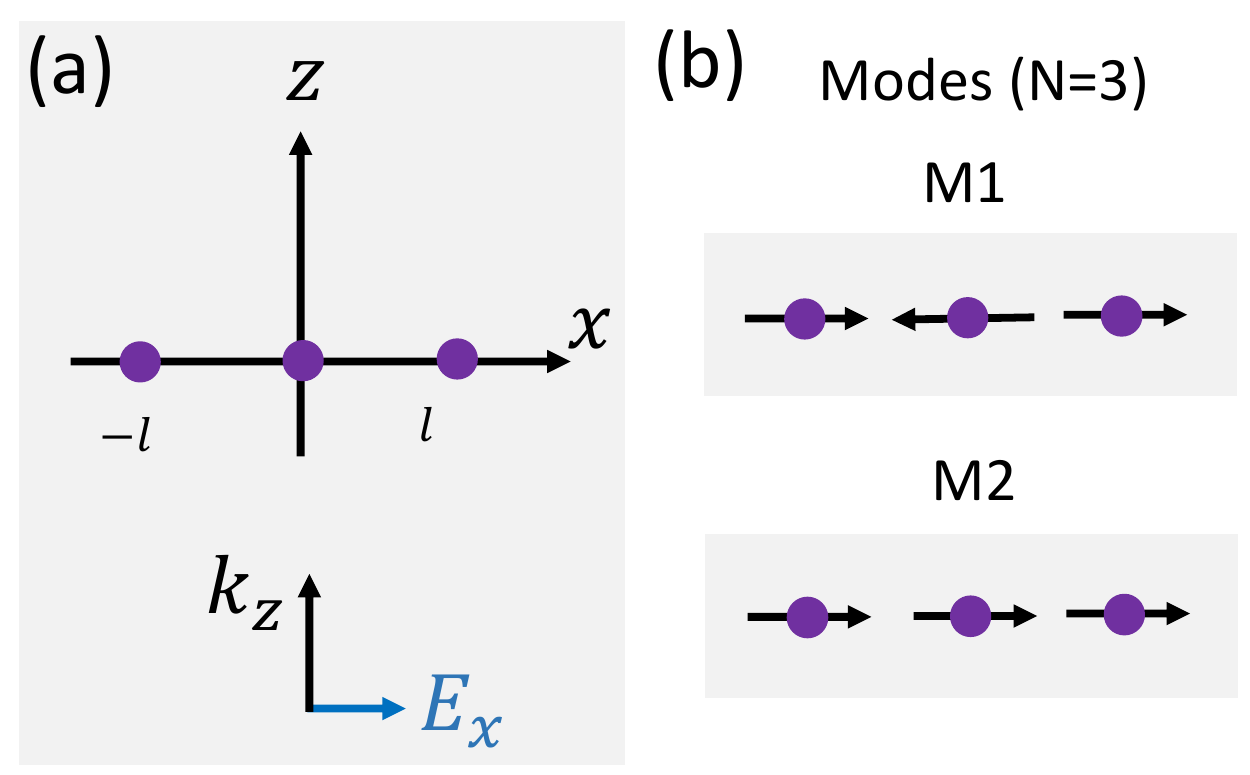}
\par\end{centering}
\caption{\textit{Scattering dark state (SDS) in an atomic trimer:} (a) Schematic
drawing of an atomic trimer placed at $\mathbf{r}_{1,3}=\pm l\mathbf{e}_{x}$,
$\mathbf{r}_{2}=0$. (b) The trimer supports two modes, i.e.,
M1 and M2 when illuminated by an $x$-polarized plane wave propagating
in the $z$ direction: $\mathbf{E}_{\mathrm{inc}}=E_{0}e^{ikz}\mathbf{e}_{x}$.
\label{fig:SDS_Trimer}}
\end{figure}

\subsection{Atomic pentamer and heptamer}
Here, we investigate scattering dark states in antennas
consisting of larger number of atoms (i.e., pentamer and heptamer) with 5 and 7 atoms, respectively. The scattering cross sections of the atomic pentamer and heptamer are plotted in Fig.~\ref{fig:SDS_tetramer_pentamer}(a)-(b), and in Fig.~\ref{fig:SDS_tetramer_pentamer}(c)-(d), respectively. It can be seen that the antenna with $n$ excited modes
exhibits $n-1$ scattering dark states~[see e.g.,  Fig.~\ref{fig:SDS_tetramer_pentamer}
(d)]. For example, the heptamer exhibits 4 modes with 3 scattering dark states with vanishing cross sections.

\begin{figure}
\begin{centering}
\includegraphics[width=10cm]{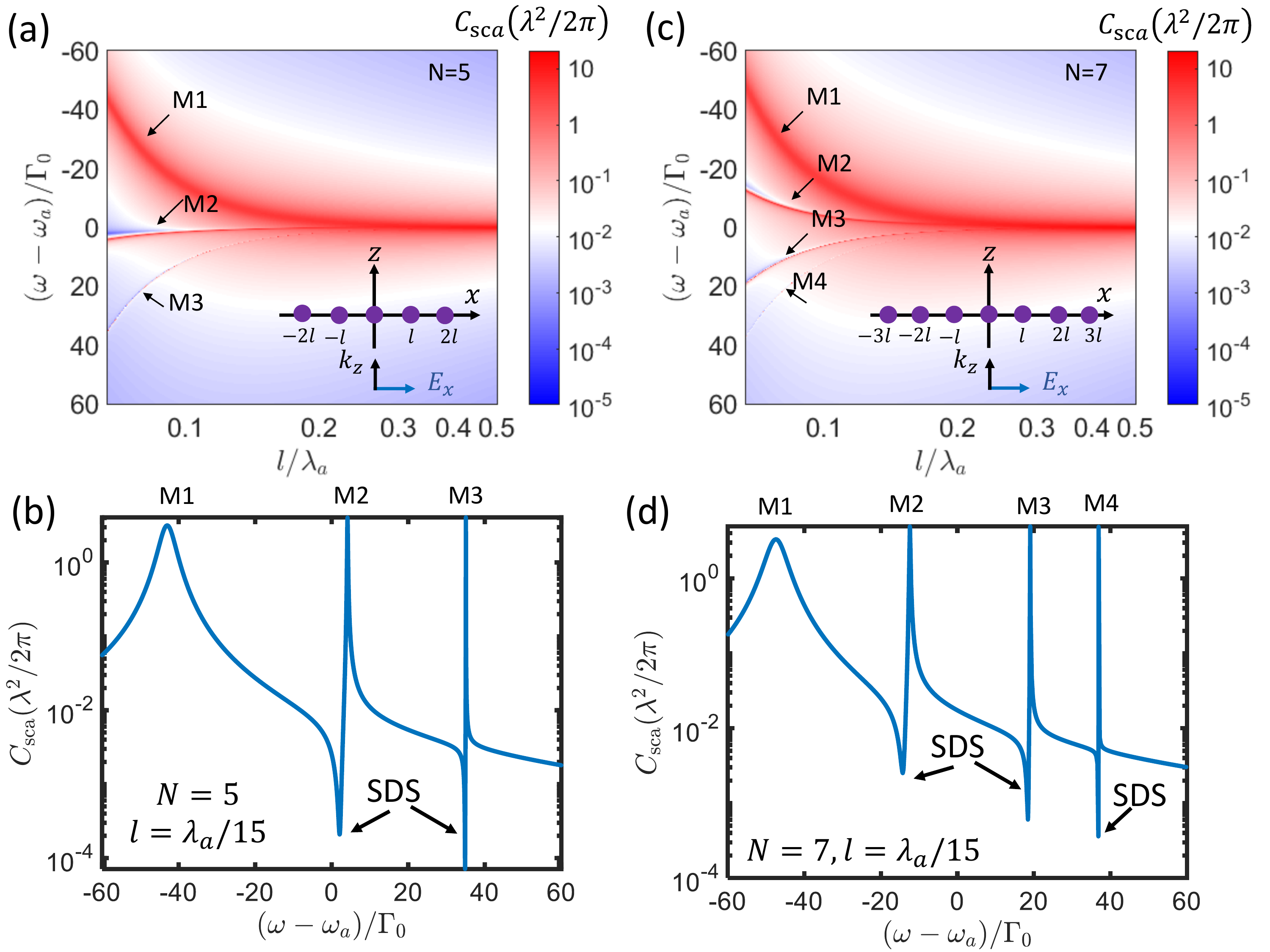}
\par\end{centering}
\caption{\textit{Scattering dark state as a nonradiating source:} (a) and (c)
Scattering cross section (normalized to $\text{\ensuremath{\lambda^{2}/2\pi}}$
and shown in logarithmic scale) as a function of frequency detuning
and the distance between the atoms $l$ for pentamer ($N=5$) and
heptamer ($N=7$), respectively. Inset: schematic drawing of the antenna
and the plane wave excitation. (b) Normalized scattering cross section
as a function of frequency detuning for $l=\lambda_{a}/15$, for pentamer
and heptamer, respectively. \label{fig:SDS_tetramer_pentamer}}
\end{figure}



\section{Kerker effect}
\subsection{Atomic trimer in an equilateral triangle configuration}
In this section, we derive Eq.~(5) of the main text, i.e., the induced electric and magnetic dipole
moments of an atomic trimer in an equilateral triangle [see Fig.~\ref{fig:Trimer_Kerker} (a)]. Here, we consider an atomic trimer in an equilateral triangle configuration
consisting of three identical atoms with an electric polarizability
$\alpha$ placed at $\mathbf{r}_{1,2}=\pm l/2\mathbf{e}_{x}-\frac{\sqrt{3}}{6}l\mathbf{e}_{z}$,
$\mathbf{r}_{3}=\frac{\sqrt{3}}{3}l\mathbf{e}_{z}$. The trimer is
illuminated by an $x$-polarized plane wave propagating in the $z$
direction~[see Fig.~\ref{fig:Trimer_Kerker}]. The induced displacement
volume current density for the trimer read as
\begin{eqnarray}
\mathbf{J}\left(\mathbf{r},\omega\right) & = & -i\omega\sum_{i=1}^{3}\mathbf{p}_{i}\left(\mathbf{r}_{i}\right)\delta\left(\mathbf{r}-\mathbf{r}_{i}\right),\label{eq:J_trimer}\\
 & = & -i\omega\left[p_{1}^{x}\delta\left(\mathbf{r}-\mathbf{r}_{1}\right)+p_{2}^{x}\delta\left(\mathbf{r}-\mathbf{r}_{2}\right)+p_{3}^{x}\delta\left(\mathbf{r}-\mathbf{r}_{3}\right)\right]\mathbf{e}_{x}-i\omega\left[p_{1}^{z}\delta\left(\mathbf{r}-\mathbf{r}_{1}\right)+p_{2}^{z}\delta\left(\mathbf{r}-\mathbf{r}_{2}\right)\right]\mathbf{e}_{z},\nonumber 
\end{eqnarray}
where $p_{1}^{x}=p_{2}^{x}$ and $p_{1}^{z}=-p_{2}^{z}$ due to the
symmetry of the trimer.
\subsection{Induced dipole moments }
First, let us start by introducing the Green function of the atomic
trimer~(shown in Fig.~\ref{fig:Trimer_Kerker})
\begin{eqnarray}
\overline{\mathbf{\overline{G}}}_{12} & = & \left[\begin{array}{ccc}
\text{\ensuremath{G_{12}^{xx}}} & 0 & 0\\
0 & \text{\ensuremath{G_{12}^{yy}}} & 0\\
0 & 0 & \text{\ensuremath{G_{12}^{zz}}}
\end{array}\right]=\frac{3}{2\alpha_{0}\varepsilon_{0}}e^{ikl}\left[\begin{array}{ccc}
g_{1}+\frac{g_{2}}{4} & 0 & 0\\
0 & g_{1} & 0\\
0 & 0 & g_{1}
\end{array}\right],\nonumber \\
\overline{\mathbf{\overline{G}}}_{13} & = & \left[\begin{array}{ccc}
\text{\ensuremath{G_{13}^{xx}}} & 0 & \text{\ensuremath{G_{13}^{xz}}}\\
0 & \text{\ensuremath{G_{13}^{yy}}} & 0\\
\text{\ensuremath{G_{13}^{xz}}} & 0 & \text{\ensuremath{G_{13}^{zz}}}
\end{array}\right]=\frac{3}{2\alpha_{0}\varepsilon_{0}}e^{ikl}\left[\begin{array}{ccc}
g_{1}+\frac{g_{2}}{4} & 0 & -\frac{\sqrt{3}g_{2}}{4}\\
0 & g_{1} & 0\\
-\frac{\sqrt{3}g_{2}}{4} & 0 & g_{1}+\frac{3g_{2}}{4}
\end{array}\right],\nonumber \\
\overline{\mathbf{\overline{G}}}_{23} & = & \left[\begin{array}{ccc}
\text{\ensuremath{G_{23}^{xx}}} & 0 & \text{\ensuremath{G_{23}^{xz}}}\\
0 & \text{\ensuremath{G_{23}^{yy}}} & 0\\
\text{\ensuremath{G_{23}^{xz}}} & 0 & \text{\ensuremath{G_{23}^{zz}}}
\end{array}\right]=\frac{3}{2\alpha_{0}\varepsilon_{0}}e^{ikl}\left[\begin{array}{ccc}
g_{1}+\frac{g_{2}}{4} & 0 & \frac{\sqrt{3}g_{2}}{4}\\
0 & g_{1} & 0\\
\frac{\sqrt{3}g_{2}}{4} & 0 & g_{1}+\frac{3g_{2}}{4}
\end{array}\right],\label{eq:GF_Kerker}
\end{eqnarray}
where $g_{1}=\left(\frac{1}{kl}-\frac{1}{\left(kl\right)^{3}}+\frac{i}{\left(kl\right)^{2}}\right),$
and $g_{2}=\left(-\frac{1}{kl}+\frac{3}{\left(kl\right)^{3}}-\frac{3i}{\left(kl\right)^{2}}\right)$.
Using the coupled dipole theory Eq.~(\ref{eq:CDT_S}) and Green's tensor,
i.e., Eq.~(\ref{eq:GF_Kerker}), we obtain the induced electric dipole
moments of each atom
\begin{eqnarray}
p_{1}^{x}	& = &	p_{2}^{x}=\varepsilon_{0}\alpha\frac{E_{0}}{D}\left[2\varepsilon_{0}^{2}\alpha^{2}G_{13}^{xz\,2}-\varepsilon_{0}\alpha\text{\ensuremath{G_{12}^{yy}}}-\varepsilon_{0}\alpha\text{\text{\ensuremath{G_{13}^{xx}}}}\left(1+\varepsilon_{0}\alpha\text{\ensuremath{G_{12}^{yy}}}\right)e^{i\frac{\sqrt{3}}{2}kl}-1\right]e^{-i\frac{\sqrt{3}}{6}kl},\nonumber \\
p_{3}^{x}	& = &	-\varepsilon_{0}\alpha\frac{E_{0}}{D}\left(1+\varepsilon_{0}\alpha\text{\ensuremath{G_{12}^{yy}}}\right)\left[2\varepsilon_{0}\alpha\text{\text{\ensuremath{G_{13}^{xx}}}}+\left(1-\varepsilon_{0}\alpha\text{\ensuremath{G_{12}^{xx}}}\right)e^{i\frac{\sqrt{3}}{2}kl}\right]e^{-ik\frac{\sqrt{3}}{6}l},\nonumber \\
p_{2}^{z} & = &	-p_{1}^{z}=\varepsilon_{0}\alpha\frac{E_{0}}{D}\left\{ \varepsilon_{0}\alpha G_{13}^{xz}\left[2\varepsilon_{0}\alpha\text{\text{\ensuremath{G_{13}^{xx}}}}+\left(1-\varepsilon_{0}\alpha\text{\ensuremath{G_{12}^{xx}}}\right)e^{i\frac{\sqrt{3}}{2}kl}\right]\right\} e^{-ik\frac{\sqrt{3}}{6}l},
\end{eqnarray}
where $D=2\varepsilon_{0}^{2}\alpha^{2}G_{13}^{xz2}\left(1-\varepsilon_{0}\alpha\text{\ensuremath{G_{12}^{xx}}}\right)-\left(1+\varepsilon_{0}\alpha\text{\ensuremath{G_{12}^{yy}}}\right)\left[1-\varepsilon_{0}\alpha\left(2G_{13}^{xx2}+\text{\ensuremath{G_{12}^{xx}}}\right)\right]$.\\

\begin{figure}
\begin{centering}
\includegraphics[width=5cm]{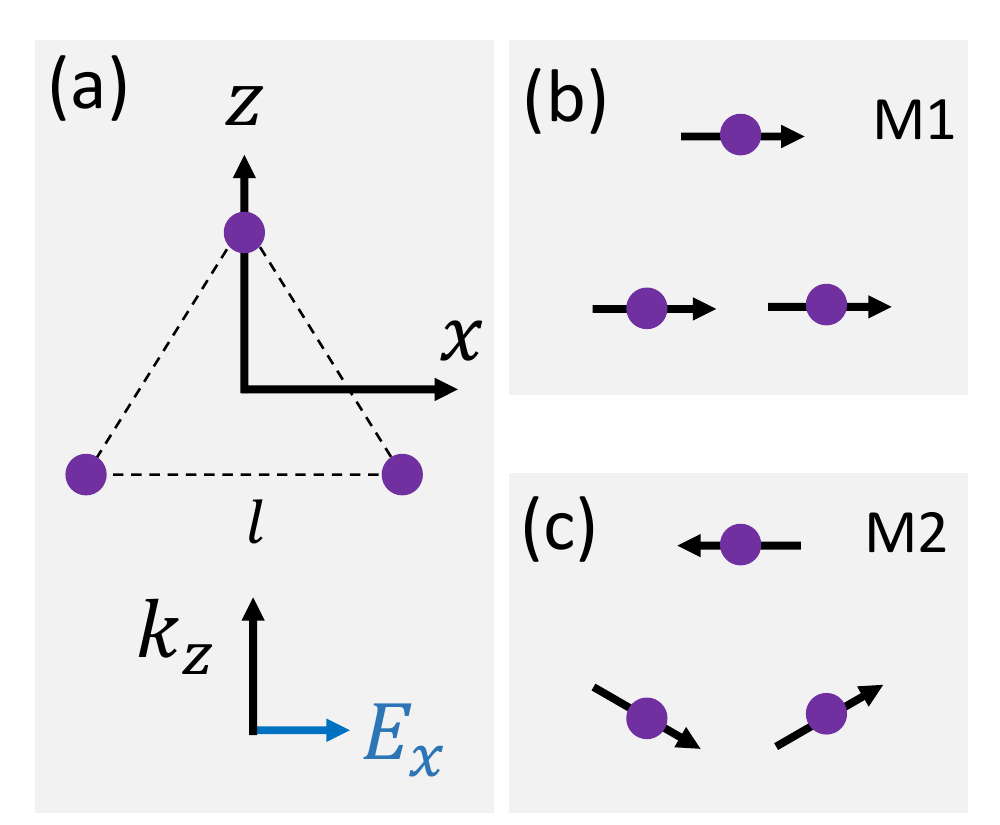}
\par\end{centering}
\caption{\textit{Kerker condition in an atomic trimer in an equilateral triangle:}
(a) Schematic drawing of of an atomic trimer in an equilateral triangle
composed of atoms at $\mathbf{r}_{1,2}=\pm l/2\mathbf{e}_{x}-\frac{\sqrt{3}}{6}l\mathbf{e}_{z}$,
$\mathbf{r}_{3}=\frac{\sqrt{3}}{3}l\mathbf{e}_{z}$ and illuminated
by an $x$-polarized plane wave propagating in the $z$ direction,
i.e., $\mathbf{E}_{\mathrm{inc}}=E_{0}e^{ikz}\mathbf{e}_{x}$, where $\mathbf{e}_{x}$
is the unit vector in the $x$ direction, $E_{0}$ is the electric
field amplitude, and $k$ is the wave vector in free space. (b)-(c) M1
and M2 are the excited modes of the trimer. \label{fig:Trimer_Kerker}}
\end{figure}
\textit{Multipole expansion at the center of the atomic trimer$-.$}
Alternatively, one can consider the trimer as an antenna. The induced \textit{effective} electric dipole moment of the trimer is calculated
by applying the multipole expansion at the center of the trimer $\mathbf{r}=0$~\cite{Alaee:2018,Alaee2019}
\begin{eqnarray}
p_{\mathrm{eff}}^{\beta} & = & -\frac{1}{i\omega}\left\{ \int dvJ_{\beta}j_{0}\left(kr\right)+\frac{k^{2}}{2}\int dv\left[3\left(\mathbf{r}\cdot\mathbf{J}\right)r_{\beta}-r^{2}J_{\beta}\right]\frac{j_{2}\left(kr\right)}{\left(kr\right)^{2}}\right\} ,\label{eq:p_ME}
\end{eqnarray}
where $\beta=x,y,z$ and $j_{n}\left(kr\right)$ are spherical Bessel
functions. $\mathbf{p}_{\mathrm{eff}}=p_{\mathrm{eff}}^{x}\mathbf{e}_{x}+p_{\mathrm{eff}}^{z}\mathbf{e}_{z}$.
Now by substituting Eq.~(\ref{eq:J_trimer}) into Eq.~(\ref{eq:p_ME}),
we get [Eq.(5) of the main text]
\begin{eqnarray}
p_{\mathrm{eff}}^{x} & = & -\frac{1}{i\omega}\left\{ \int J_{x}j_{0}\left(kr\right)dv+\frac{k^{2}}{2}\int dv\left[3\left(\mathbf{r}\cdot\mathbf{J}\right)x-r^{2}J_{x}\right]\frac{j_{2}\left(kr\right)}{\left(kr\right)^{2}}\right\} ,\nonumber \\
 & = & -\frac{1}{i\omega}\int J_{x}\left[j_{0}\left(kr\right)+\frac{1}{2}\left(3\frac{x^{2}}{r^{2}}-1\right)j_{2}\left(kr\right)\right]dv,\nonumber \\
 & = & \left[p_{1}^{x}+p_{2}^{x}+p_{3}^{x}\right]j_{0}\left(\frac{\sqrt{3}}{3}kl\right)+\frac{1}{8}\left[5p_{1}^{x}+5p_{2}^{x}-4p_{3}^{x}\right]j_{2}\left(\frac{\sqrt{3}}{3}kl\right),\nonumber \\
 & = & \left[2p_{1}^{x}+p_{3}^{x}\right]j_{0}\left(\frac{\sqrt{3}}{3}kl\right)+\frac{1}{4}\left[5p_{1}^{x}-2p_{3}^{x}\right]j_{2}\left(\frac{\sqrt{3}}{3}kl\right).\label{eq:px_Kerker}
\end{eqnarray}
In Eq.~\ref{eq:px_Kerker}, we used $p_{1}^{x}=p_{2}^{x}$. Using Eq.~(\ref{eq:J_trimer}) and $\,p_{1}^{z}=-p_{2}^{z},$ we obtion
\begin{eqnarray}
p_{\mathrm{eff}}^{z} & = & -\frac{1}{i\omega}\left\{ \int J_{z}j_{0}\left(kr\right)dv+\frac{k^{2}}{2}\int dv\left[3\left(\mathbf{r}\cdot\mathbf{J}\right)z-r^{2}J_{z}\right]\frac{j_{2}\left(kr\right)}{\left(kr\right)^{2}}\right\} ,\nonumber \\
 & = & -\frac{1}{i\omega}\int J_{z}\left[j_{0}\left(kr\right)+\frac{1}{2}\left(3\frac{z^{2}}{r^{2}}-1\right)j_{2}\left(kr\right)\right]dv=0.
\end{eqnarray}
Note that the $y$-component of the electric dipole moment is also zero,
i.e. $p_{\mathrm{eff}}^{y}=0$. 

The induced \textit{effective} magnetic
dipole moment at the center of the trimer $\mathbf{r}=0$ reads as~\cite{Alaee:2018,Alaee2019}
\begin{eqnarray}
m_{\mathrm{eff}}^{\beta} & = & \frac{3}{2}\int dv\left(\mathbf{r}\times\mathbf{J}\right)_{\beta}\frac{j_{1}\left(kr\right)}{kr},\label{eq:m_ME}
\end{eqnarray}
where $\beta=x,y,z$. By substituting Eq.~\ref{eq:J_trimer} into
Eq.~\ref{eq:m_ME}, we obtain
\begin{eqnarray}
m_{\mathrm{eff}}^{y} & = & \frac{3}{2}\int dv\left(\mathbf{r}\times\mathbf{J}\right)_{y}\frac{j_{1}\left(kr\right)}{kr}=\frac{3}{2k}\int\left(\frac{z}{r}J_{x}-\frac{x}{r}J_{z}\right)j_{1}\left(kr\right)dv,\nonumber \\
 & = & i\frac{3}{2}c\left[\frac{p_{1}^{x}}{2}+\frac{p_{2}^{x}}{2}-p_{3}^{x}\right]j_{1}\left(\frac{\sqrt{3}}{3}kl\right)+i\frac{3\sqrt{3}}{4}c\left[p_{2}^{z}-p_{1}^{z}\right]j_{2}\left(\frac{\sqrt{3}}{3}kl\right),\nonumber \\
  & = & \frac{3i}{2}c\left[\left(p_{1}^{x}-p_{3}^{x}\right)j_{1}\left(u\right)+\sqrt{3}p_{2}^{z}j_{2}\left(u\right)\right]\label{eq:my_Kerker}
\end{eqnarray}
, which is Eq. (4) of the main text.
Note that one can show that $x$- and $z$- components of the magnetic
dipole moments are zero, i.e., $m_{\mathrm{eff}}^{x}=0$, and $m_{\mathrm{eff}}^{z}=0$.
The induced multipole moments of the trimer in Fig.~5 of the main
text are calculated using Eq.~(\ref{eq:px_Kerker}) and Eq.~(\ref{eq:my_Kerker}). 


\subsection{Radiation pattern}
Having the induced multipole moments of the trimer, i.e., Eq.~(\ref{eq:px_Kerker}) and Eq.~(\ref{eq:my_Kerker}), the far field of an atomic antennas can be found using~\cite{Jackson1999,Alaee_kerker:15}
\begin{eqnarray}
\mathbf{E}_{\rm ED} & = & \frac{k^{2}}{4\pi\epsilon_{0}}\frac{e^{ikr}}{r}p_{x}\left(-\mathrm{sin}\varphi\mathbf{e}_{\varphi}+\mathrm{cos}\theta\mathrm{cos}\varphi\mathbf{e}_{\theta}\right),\nonumber \\
\mathbf{E}_{\rm MD} & = & \frac{k^{2}}{4\pi\epsilon_{0}}\frac{e^{ikr}}{r}\frac{m_{y}}{c}\left(-\mathrm{cos}\theta\mathrm{sin}\varphi\mathbf{e}_{\varphi}+\mathrm{cos}\varphi\mathbf{e}_{\theta}\right),\nonumber \\
\mathbf{E}_{\rm EQ} & = & \frac{k^{2}}{4\pi\epsilon_{0}}\frac{e^{ikr}}{r}\frac{ik}{6}Q_{zx}^{e}\left[\mathrm{cos}\theta\mathrm{sin}\varphi\mathbf{e}_{\varphi}-\mathrm{\left(\mathrm{2cos^{2}}\theta-1\right)cos}\varphi\mathbf{e}_{\theta}\right],\nonumber \\
\mathbf{E}_{\rm MQ} & = & \frac{k^{2}}{4\pi\epsilon_{0}}\frac{e^{ikr}}{r}\frac{ik}{6c}Q_{zy}^{m}\left[\left(\mathrm{2cos^{2}}\theta-1\right)\mathrm{sin}\varphi\mathbf{e}_{\varphi}-\mathrm{\mathrm{cos}\theta cos}\varphi\mathbf{e}_{\theta}\right],\label{eq:E_Far}
\end{eqnarray}
where $r,\theta,\varphi$ are the radial distance, polar angle, and
azimuthal angle, respectively. The radiation pattern ($\propto\left|\mathbf{E}\right|^{2}$) in $xz$-plane, i.e., $\varphi=0$, considering the contribution from all multipole moments (up
to magnetic quadrupole), can be written as
\begin{eqnarray}
\mathbf{E} & \approx & \frac{k^{2}}{4\pi\epsilon_{0}}\frac{e^{ikr}}{r}\left[p_{x}\mathrm{cos}\theta+\frac{m_{y}}{c}-\frac{ik}{6}Q_{xz}^{e}\left(\mathrm{2cos^{2}}\theta-1\right)-\frac{ik}{6c}Q_{zy}^{m}\mathrm{cos}\theta\right]\mathbf{e}_{\theta}.
\end{eqnarray}
For an atomic trimer with $\lambda\gg l$ at the M2 mode, i.e. $\omega_{\mathrm{K}}$,
the electric and magnetic quadruple moments are negligible, i.e. $Q_{xz}^{e}\approx0$,
and $Q_{yz}^{m}\approx0$. Thus, the radiation pattern ($\propto\left|\mathbf{E}\right|^{2}$) is mainly due
to electric and magnetic dipole moments, i.e.,
\begin{eqnarray}
\mathbf{E} & \approx & \frac{k^{2}}{4\pi\epsilon_{0}}\frac{e^{ikr}}{r}\left[p_{x}\mathrm{cos}\theta+\frac{m_{y}}{c}\right]\mathbf{e}_{\theta},
\end{eqnarray}
where $p_{x}$ and $m_{y}$ for the atomic trimer in Fig.~\ref{fig:Trimer_Kerker}
can be obtained using Eq.~(\ref{eq:px_Kerker}) and (\ref{eq:my_Kerker}),
respectively.

\end{widetext}

%

\end{document}